\documentclass[letterpaper, 10 pt]{IEEEtran}  

\IEEEoverridecommandlockouts

\usepackage{amsmath} 
\usepackage{amssymb} 

\usepackage{amsthm}
\usepackage{graphicx}
\usepackage[space]{grffile}
\usepackage{cite}
\usepackage{wrapfig}

\expandafter\def\expandafter\normalsize\expandafter{%
    \normalsize
    \setlength\abovedisplayskip{3pt}
    \setlength\belowdisplayskip{4pt}
    \setlength\abovedisplayshortskip{3pt}
    \setlength\belowdisplayshortskip{3pt}
}

\title{\LARGE \bf
Stabilizing Value Iteration with and without Approximation Errors
}

\author{Ali Heydari$^1$
\thanks{$^{1}$Assistant Professor of Mechanical Engineering, South Dakota School of Mines and Technology, Rapid City, SD 57701, email: ali.heydari@sdsmt.edu.}}

\newtheorem{Thm}{Theorem} 
\newtheorem{Lem}{Lemma} 
 
\newtheorem{Def}{Definition} 
\newtheorem{Assumption}{Assumption} 

\begin{document}

\maketitle
\pagenumbering{arabic}
\pagestyle{plain}
\thispagestyle{plain}

\begin{abstract}
Adaptive optimal control using value iteration (VI) initiated from a stabilizing policy is theoretically analyzed in various aspects including the continuity of the result, the stability of the system operated using any \textit{single/constant} resulting control policy, the stability of the system operated using the \textit{evolving/time-varying} control policy, the convergence of the algorithm, and the optimality of the limit function. Afterwards, the effect of presence of approximation errors in the involved function approximation processes is incorporated and another set of results for boundedness of the approximate VI as well as stability of the system operated under the results for both cases of applying a single policy or an evolving policy are derived. A feature of the presented results is providing estimations of the \textit{region of attraction} so that if the initial condition is within the region, the whole trajectory will remain inside it and hence, the function approximation results will be reliable. 

\end{abstract}
\section{Introduction}
Intelligent control using adaptive/approximate dynamic programming (ADP), sometimes referred to by reinforcement learning (RL) or neuro-dynamic programming (NDP), is a set of powerful tools for obtaining approximate solutions to difficult and mathematically intractable problems which seek optimum while sometimes even no knowledge of the system model/dynamics is available. The dramatic potential of the tools in practice has attracted many researchers within the last few decades, \cite{Watkins}-\cite{Heydari_NN_FinHor}. 
The multitude of appeared papers and success stories on applications of ADP to different problems, however, has intensified the need for firm mathematical analyses for guaranteeing the convergence of the learning processes and the stability of the results. 

Besides the classifications of heuristic dynamic programming (HDP), dual heuristic programming (DHP), etc. \cite{Prokhorov}, which are in terms of the variables subject to approximation and their dependencies, the learning algorithms are typically based on either value iteration (VI) or policy iteration (PI), \cite{Sutton, LewisContSystMag}. These algorithms are well investigated both by computer scientists for machine learning \cite{Sutton} and by control scientists for feedback control of dynamical systems \cite{LewisContSystMag}. PI, despite having a higher computational load due to a `full backup' as opposed to a `partial backup' in VI \cite{LewisContSystMag}, has the advantage that the control under evolution remains stabilizing, \cite{Liu_PI}. Hence, PI seems more suitable for online implementation, i.e., adapting the control `on the fly'. However, the requirement that PI needs to start with an stabilizing initial control is one of its drawbacks. VI, on the other hand, does not require an stabilizing initial control and can be initiated arbitrarily. But, the closed loop system is not guaranteed to be stable during its learning process, if implemented online.

Considering optimal control of discrete-time problems with continuous state and action spaces and undiscounted cost functions using VI, which is the subject of this work, the convergence proof for linear systems was analyzed in \cite{Landelius_PhDThesis, Bala_Liu_Convergence}. As for nonlinear systems, the convergence was established by different researchers including \cite{Lincol_RelaxingDynProg} (adapted in \cite{Rinehart_VI_TAC}), \cite{AlTamimi}, and \cite{Heydari_TCYB} through different approaches. 
All these convergence analyses are based on the assumption of perfect function reconstruction, i.e., no error in the function approximation. 
While this assumption plays a major rule in deriving the results, it restricts their practical use severely, because, the approximation errors exist almost in every application when the system is nonlinear or when the cost function terms are non-quadratic and nonlinear. What makes their presence potentially problematic is the fact that the errors \textit{propagate} throughout the iterations, hence, regardless of how small they are, a phenomenon similar to \textit{resonance} might happen which could lead to the complete unreliability of the results. 

Analyzing VI under the presence of approximation errors, i.e. \textit{approximate VI} (\textit{AVI}), is an open research problem with a few published results, including \cite{Bertsekas_NDP, Singh_Discounted_ADP_ErrorAnalysis, Szepesv_AVI_API_ErrorAnalysis, Szepesv_AVI_ErrorAnalysis, Liu_TCYB}, to the best of the knowledge of the author. Refs. \cite{Bertsekas_NDP, Singh_Discounted_ADP_ErrorAnalysis, Szepesv_AVI_API_ErrorAnalysis, Szepesv_AVI_ErrorAnalysis} investigated problems with \textit{discounted} cost functions and the results are solely valid for such problems, prevalent in computer science. As a matter of fact, the `forgetting' nature of discounted problems is the backbone of the developments of the error bounds and if the discount factor approaches one, as in typical infinite-horizon optimal control problems, the bounds go to infinity. Hence, the results do not cover this case.
On the other hand, the interesting results in \cite{Liu_TCYB} provide some error analyses but with assumptions which are more restrictive and not easily verifiable, compared with this study. For example, the approximation error between the exact and approximate functions, respectively denoted with $V(.)$ and $\hat{V}(.)$, should be possible to be written in the \textit{multiplicative} form of $\hat{V}(x) \leq \sigma V(x)$ for some positive constant $\sigma, \forall x$, instead of an \textit{additive} form of $\hat{V}(x) = V(x) + \epsilon(x)$, for some real valued function $\epsilon(.)$. 
Moreover, the boundedness results are conditional upon $\sigma$ being upper bounded with a term including a parameter which corresponds to the \textit{optimal} value function. 

Based on this background, besides the stability issue during the online learning stage using VI, rigorous theoretical analyses of the consequences of the errors on the results are of great interest to the ADP researchers and practitioners. The reasons are the scarcity of the available studies on AVI, the prevalence of approximation errors, and the great potential of the tool in (approximately) solving optimal control problems in practice.

The contributions of this study are multiple. Initially, it is proved that VI also will be stabilizing for online control if, similar to PI, it is started using an initial stabilizing control. 
Afterwards, it is shown that the start from an initial stabilizing control leads to an initial value function that does not satisfy the necessary conditions for any of the cited convergence proofs of VI. Establishing this convergence (to the optimal solution) is another contribution of this work. 
These results may not look substantially different from what the ADP community assumes to hold intuitively or has already established, \cite{Liu_Stable_VI}.
What makes the abovementioned two results different is having two characteristics. The theoretical analyses in this study are simple and straight forward, both for optimality and stability analyses, compared to the existing developments in the literature. Another feature is providing rigorous mathematical bases for the analyses. As an example, the use of a value function as a Lyapunov function for stability analysis requires proof of continuity of the value function, \cite{Khalil}. The firm proof of this continuity, presented in this work, is not as straight forward as it looks. The factors leading to the difficulty are the presence of the $argmin$ operator in calculation of the control at each iteration of VI, which may potentially lead to a discontinuity in the control policy, and also the concern of \textit{pointwise} versus \textit{uniform} convergence, for concluding the continuity of the limit function from the continuity of the elements of a converging sequence of functions, \cite{Rudin}, \cite{Rinehart_VI_TAC}. 

Another contribution of this work is addressing the legitimate concern that any ADP result is valid only when the state trajectory remains within the domain for which the controller is trained. 
This concern is resolved through establishing an \textit{estimation of the region of attraction} (EROA) \cite{Khalil} for the controller, in this work, so that as long as the system's initial condition is within the region, it is guaranteed that the entire trajectory remains in the region. Hence, the controller will remain valid and usable.

As the reader delves into the problem, it is discussed that the provided stability proof, whose main idea is not much different from \cite{Liu_Stable_VI}, assumes applying a fixed (time-invariant) control policy on the system. But, this is rarely the case in online learning, since, as the learning proceeds, the control policy \textit{evolves}, hence, the applied control policy is time-varying. 
Therefore, another set of stability results for the time-varying and evolving control policy is developed with some ideas for establishing its respective EROA, as another contribution of this work. 

After providing the detailed analysis of the VI which is initiated with an admissible guess, called Stabilizing VI throughout the paper, the case of presence of the approximation errors is investigated, leading to a new set of contributions which are of greater interests. They include boundedness/convergence analysis of the AVI initiated with an admissible guess, stability and EROA analysis for the case of applying a fixed control policy, and stability and EROA analysis for the case of applying an evolving control policy. These theoretical analyses are the most important contributions of this work, as the assumptions leading to the results are verifiable and more straight forward, compared with the available studies.
Finally, interested readers are referred to \cite{Heydari_AVI} for some recent developments of this author on analyzing the effect of the approximation errors in regular value iteration, i.e., the approximate VI which is initiated arbitrarily. \footnote{It must be added that the current version of this paper has overlaps with the first version of \cite{Heydari_AVI} on Lemma \ref{Lemma_Cont_W_vs_V} and Theorem \ref{Thm_Boundedness}.}
 
The rest of this paper is organized as follows. The problem is formulated in Section II and the ADP-based solutions are revisited in Section III. Section IV presents the theoretical analyses on exact VI. The respective analyses for AVI are presented in section V. 
Finally, concluding remarks are given in Section VI.

\section{Problem Formulation} \label{ProblemFormulation}
Let the system subject to control be given by discrete-time nonlinear dynamics
\begin{equation}
x_{k+1}=f(x_k,u_k), k \in \mathbb{N}, \label{Dynamics}
\end{equation}
where $f:\mathbb{R}^n \times \mathbb{R}^m \to \mathbb{R}^n$ is a Lipschitz continuous function versus its both inputs, i.e., the state and control vectors, $x$ and $u$, respectively, with $f(0,0)=0$. The set of non-negative integers is denoted with $\mathbb{N}$, and positive integers $n$ and $m$ denote the dimensions of the continuous state and control spaces. Finally, sub-index $k$ represents the discrete time index. 
The performance index is given by 
\begin{equation}
J=\sum_{k=0}^\infty {U(x_k,u_k)}, \label{CostFunction}
\end{equation}
where \textit{utility function} $U(.,.)$ is of form $U(x_k,u_k):=Q(x_k) + u_k^TRu_k$ for a continuous and positive semi-definite function $Q:\mathbb{R}^n \to \mathbb{R}_+$ and a positive definite $m \times m$ real matrix $R$. Set $\mathbb{R}_+$ denotes the non-negative reals. 
Starting with any initial feedback \textit{control policy} given by $h:\mathbb{R}^n \to \mathbb{R}^m$ for control calculation, i.e., $u_k = h(x_k)$, the problem is \textit{updating/adapting} the control policy such that cost function (\ref{CostFunction}) is minimized. The control policy which leads to such a characteristic is called \textit{optimal control policy}, denoted with $h^*(.)$.

\begin{Def} \label{AsymStability_Definition}
A control policy is defined to be {asymptotically stabilizing} within a domain if $lim_{k\to\infty} x_k = 0$ using this control policy, for every initial state within the domain, \cite{Khalil}.   
\end{Def} 

\begin{Def} \label{Def1}
A control policy $h(.)$ is defined to be \textit{admissible} within a compact set if a) it is a Lipschitz continuous function of $x$ in the set with $h(0)=0$, b) it asymptotically stabilizes the system within the set, and 
c) there exists a continuous positive definite function $W:\mathbb{R}^n \to \mathbb{R}_+$ that puts an upper bound on the respective \textit{`cost-to-go'} or \textit{`value function'}, denoted with $V_h:\mathbb{R}^n \to \mathbb{R}_+$ and defined by 
\begin{equation}
V_h(x_0)=\sum_{{k}=0}^\infty {U\big(x_{k}^h, h(x_{k}^h)\big)}, \label{ValueFunction_of_h}
\end{equation}
i.e., $V_h(x) \leq W(x), \forall x \in \Omega$. 
In Eq. (\ref{ValueFunction_of_h}) one has $x_k^h:=f\big(x_{k-1}^h,h(x_{k-1}^h)\big), \forall k \in \mathbb{N}-\{0\},$ and $x_0^h := x_0$. In other words, $x_k^h$ denotes the $k$th element on the state trajectory/history initiated from $x_0$ and propagated using control policy $h(.)$.
\end{Def}

The main difference between the defined admissibility and the ones typically utilized in the ADP/RL literature, including \cite{AlTamimi}, is the assumption of upper boundedness of the value function by a continuous (positive definite) function. This condition is trivially satisfied if the value function itself is continuous, i.e., through selecting $W(.)=V_h(.)$. However, instead of assuming continuity of the value function, the milder condition of being upper bounded by such a function is assumed. Note that the continuity of the value function is required for \textit{uniform} approximation of the function using parametric function approximators, \cite{Weierstrass_Theorem, Hornik_NN_Continuity}. In this study, it will be shown that the upper boundedness will lead to the desired continuity. Finally, it should be noted that continuous functions are bounded in a compact set \cite{Rudin}, hence, the upper boundedness of the value function by the continuous function $W(.)$ leads to the boundedness of the respective value function. This is an essential requirement for an admissible control, as a mere asymptotically stabilizing control policy may lead to an unbounded value function.

\begin{Assumption} \label{Assum_ExistingAdmissibleCont}
There exists at least one admissible control policy for the given system within a connected and compact set $\Omega \subset \mathbb{R}^n$ containing the origin.
\end{Assumption}

\begin{Assumption} \label{Assum_InvariantSet}
The intersection of the set of n-vectors $x$ at which $U(x,0) = 0$ with the invariant set of $f(.,0)$ only contains the origin. 
\end{Assumption}

Assumption \ref{Assum_ExistingAdmissibleCont} guarantees that there is no state vector in $\Omega$ for which the value function associated with the \textit{optimal} control policy is infinite. 
Assumption \ref{Assum_InvariantSet} assures that there is no set of states (besides the set containing only the origin) in which the state trajectory can \textit{hide} forever, in the sense that the utility function evaluated at those states is zero without convergence of the states to the origin. Note that, if such a set exists, then starting from an initial state within the set, the optimal solution would be $u_k = 0, \forall k$.

\section{ADP-based Solutions}
Based on Eq. (\ref{ValueFunction_of_h}), it can be seen that the value function satisfies the recursive relation given by
\begin{equation}
V_h(x)= U\big(x,h(x)\big) + V_h\Big(f\big(x,h(x)\big)\Big), \forall x \in \mathbb{R}^n. \label{Recursive_CostToGo}
\end{equation}
Defining the \textit{optimal value function}, as the value function associated with the optimal control policy and denoting it with $V^*(.)$, the Bellman equation \cite{Kirk}, given below, provides the solution to the problem
\begin{equation}
		h^*(x) = argmin_{u\in\mathbb{R}^m} \Big( U\big(x,u\big) + V^*\big(f\big(x,u\big)\big)\Big), \label{Bellman_eq2}
\end{equation}
\begin{equation}
		V^*(x) = min_{u\in\mathbb{R}^m} \Big( U\big(x,u\big) + V^*\big(f\big(x,u\big)\big)\Big). \label{Bellman_eq1}
\end{equation}
Due to the \textit{curse of dimensionality} \cite{Kirk}, however, the proposed solution is mathematically impracticable for general nonlinear systems. ADP utilizes the idea of \textit{approximating} the optimal value function, using either look-up tables or function approximators, e.g., neural networks (NNs), for remedying the problem. The value function approximator is typically called the \textit{critic} in the ADP/RL literature. The approximation is performed over a \textit{compact} and \textit{connected} set containing the origin, called the \textit{domain of interest}. This domain, denoted with $\Omega$, has to be selected based on the specific problem at hand and it should be noted that the ADP based results are valid only if the entire state trajectory initiated from the initial state vector remains within the domain for which the value function is approximated. 
The optimal value function approximation process is typically done through PI or VI. 
In PI, one starts with an initial admissible control policy, denoted with $h^0(.)$, and iterates through the \textit{policy evaluation equation} given by
\begin{equation}
		V^i(x) = U\big(x,h^i(x)\big) + V^i\Big(f\big(x,h^i(x)\big)\Big), \forall x \in \Omega, \label{PI_PolicyEval}
\end{equation}
and the \textit{policy update equation} given by
\begin{equation}
		h^{i+1}(x) = argmin_{u\in\mathbb{R}^m} \Big( U\big(x,u\big) + V^i\big(f\big(x,u\big)\big)\Big), \forall x \in \Omega, \label{PI_PolicyUpdate}
\end{equation}
for $i=0,1,...$ until the parameters converge. In other words, starting from policy $h^0(.)$, pointwise values for approximating value function $V^0(.)$ can be calculated using (\ref{PI_PolicyEval}) and once $V^0(.)$ is used in (\ref{PI_PolicyUpdate}) pointwise values for approximating $h^1(.)$ can be calculated, and so on.

On the other hand, in VI, the iterative learning starts with an initial guess $V^0(.)$ and iterates through the \textit{policy update equation} given by
\begin{equation}
		h^{i}(x) = argmin_{u\in\mathbb{R}^m} \Big( U\big(x,u\big) + V^i\big(f\big(x,u\big)\big)\Big), \forall x \in \Omega, \label{VI_PolicyUpdate}
\end{equation}
and the \textit{value update equation} 
\begin{equation}
		V^{i+1}(x) = U\big(x,h^i(x)\big) + V^i\Big(f\big(x,h^i(x)\big)\Big), \forall x \in \Omega, \label{VI_ValueUpdate2}
\end{equation}
or equivalently 
\begin{equation}
		V^{i+1}(x) = min_{u\in\mathbb{R}^m} \Big( U\big(x,u\big) + V^i\big(f\big(x,u\big)\big)\Big), \forall x \in \Omega, \label{VI_ValueUpdate}
\end{equation}
for $i=0,1,...$ until the iterations converge. 

\section{Analysis of Stabilizing Value Iteration}

PI requires an admissible policy $h^0(.)$ to start the process with, otherwise, there may not exist a bounded (and continuous) function $V^0(.)$ which satisfies Eq. (\ref{PI_PolicyEval}) for $i=0$. This can be observed by realizing that $V^0(.)$ is actually the value function associated with policy $h^0(.)$. 
To confirm this one may compare Eq. (\ref{PI_PolicyEval}) with Eqs. (\ref{Recursive_CostToGo}) and (\ref{ValueFunction_of_h}).

An important feature of PI is the fact that $V^i(.)$s remain Lyapunov functions for the closed loop system, hence, each respective control policy will be stabilizing. This can be confirmed by noting that from Eq. (\ref{PI_PolicyEval}) one has
\begin{equation}
		\begin{split}
		\Delta V^i(x) := V^i\Big(f\big(x,&h^i(x)\big)\Big) - V^i(x) = \\
		&- U\big(x,h^i(x)\big) \leq 0, \forall x \in \Omega. \label{PI_PolicyEval_Lyap}
		\end{split}
\end{equation}
This feature leads to its suitability for \textit{online} implementation, i.e., for \textit{adaptive} optimal control. It is an important point that \textit{online} learning has the advantage of not requiring perfect knowledge of the internal dynamics of the system \cite{AlTamimi}. However, stability of the system operated using the evolving control is of critical importance.

The VI scheme, however, can be arbitrarily initiated using any $V^0(.)$. The convergence to the optimal solution is proved for
$V^0(x)=0, \forall x,$ in \cite{AlTamimi} and for any smooth $V^0(.)$ which satisfies $0 \leq V^0(x) \leq U(x,0), \forall x,$ in \cite{Heydari_TCYB}. 
Moreover, the convergence is proved in \cite{Lincol_RelaxingDynProg} for $\alpha V^*(x) \leq V^0(x) \leq V^*(x), \alpha \in [0,1], \forall x,$ assuming there exists a finite $\beta$, independent of $x$, such that $V^*\big(f(x,u)\big) \leq \beta U(x,u), \forall x$. 

VI is more interesting for control of nonlinear systems as the convergence to optimal solution is guaranteed without requiring an initial admissible control, and also, instead of solving equation (\ref{PI_PolicyEval}), known as a `full backup', the `partial backup' given by the simple recursion (\ref{VI_ValueUpdate2}) is needed, \cite{LewisContSystMag}. However, these advantages come with the disadvantage that the `immature' control, i.e., $h^i(.)$s before the convergence of the solution, are not guaranteed to be stabilizing. This issue is generally considered as a shortcoming of VI as opposed to PI in the ADP literature, see \cite{Liu_PI} as an example. In this study, it will be shown that the credit for stabilizing feature of $h^i(.)$s in PI is due to the initial admissible control and if VI also is initiated with such a control policy, the resulting control policies under iterations will be stabilizing.

Let the initial guess, $V^0(.)$, to be used in VI, be given by the value function of an admissible control policy. For the sake of brevity, such a VI is called \textit{stabilizing value iteration}, throughout the paper, as defined in the next definition.

\begin{Def} \label{StabilizingVI_Definition}
The value iteration scheme given by recursive relation (\ref{VI_ValueUpdate}) which is initiated using the value function of an admissible control policy is called stabilizing value iteration.
\end{Def} 

Denoting the initial admissible policy with $h^{-1}(.)$ (for notational compatibility), its value function, denoted with $V^0(.)$, can be calculated through solving
\begin{equation}
		V^0(x) = U\big(x,h^{-1}(x)\big) + V^0\Big(f\big(x,h^{-1}(x)\big)\Big), \forall x \in \Omega, \label{VI_Value_eq1}
\end{equation}
per (\ref{Recursive_CostToGo}). 
Utilizing this $V^0(.)$ in the VI as the initial guess, the stability of $h^i(.)$s can be guaranteed, as proved here, which requires some theoretical results given next.

\begin{Lem} \label{Lemma_NonDecreasing} 
Sequence of functions $\{ V^{j}(x) \}_{j=0}^{\infty} := \{V^{0}(x), V^{1}(x), ... \}$ resulting from stabilizing value iteration is a pointwise non-increasing sequence.
\end{Lem}
\textit{Proof}: 
The proof is done by induction. Considering (\ref{VI_Value_eq1}), which gives $V^0(.)$, and (\ref{VI_ValueUpdate}), which (for $i=0$) gives $V^1(.)$, one has 
\begin{equation}
		V^1(x) \leq V^0(x), \forall x \in \Omega, \label{Lemma1_eq1}
\end{equation}
because $V^1(.)$ is the result of minimization of the right hand side of (\ref{VI_Value_eq1}) instead of being resulted from a given $h^{-1}(.)$.
Now, assume that for some $i$, we have
\begin{equation}
		V^i(x) \leq V^{i-1}(x), \forall x \in \Omega. \label{Lemma1_eq2}
\end{equation}
Define $\mathcal{V}(.)$ as
\begin{equation}
		\mathcal{V}(x) := U\big(x,h^{i-1}(x)\big) + V^i\Big(f\big(x,h^{i-1}(x)\big)\big), \forall x \in \Omega. \label{Lemma1_eq3}
\end{equation}
Comparing (\ref{Lemma1_eq3}) with (\ref{VI_ValueUpdate}), one has 
\begin{equation}
		V^{i+1}(x) \leq \mathcal{V}(x), \forall x \in \Omega, \label{Lemma1_eq4}
\end{equation}
because $V^{i+1}(.)$ is the result of minimization of the right hand side of (\ref{Lemma1_eq3}). Moreover, $V^{i}(.)$ is given by
\begin{equation}
		V^{i}(x) = U\big(x,h^{i-1}(x)\big) + V^{i-1}\Big(f\big(x,h^{i-1}(x)\big)\Big), \forall x \in \Omega, \label{Lemma1_eq5}
\end{equation} 
based on (\ref{VI_ValueUpdate2}). Hence, 
\begin{equation}
		\mathcal{V}(x) \leq V^{i}(x), \forall x \in \Omega, \label{Lemma1_eq6}
\end{equation}
because of (\ref{Lemma1_eq2}). 
Considering (\ref{Lemma1_eq4}) and (\ref{Lemma1_eq6}), one has 
\begin{equation}
		V^{i+1}(x) \leq V^{i}(x), \forall x \in \Omega, \label{Lemma1_eq7}
\end{equation}
which together with (\ref{Lemma1_eq2}), proves the lemma, by induction.
\qed

Before proceeding to the stability theorem, it is needed to analyze the continuity of each $V^i(.)$ within $\Omega$. Note that even though functions $f(.,.)$ and $U(.,.)$ are assumed to be continuous versus all the inputs, the existence of $argmin$ operator in Eq. (\ref{VI_PolicyUpdate}) may lead to discontinuity in function $h^i(.)$, which may then lead to a discontinuous $V^{i+1}(.)$ in Eq. (\ref{VI_ValueUpdate2}). 
Besides the $argmin$ issue, does the continuity of $h^{-1}(.)$ lead to the continuity of its value function, $V^0(.)$? Note that $V^0(.)$ is going to be the initial guess in the stabilizing VI, hence, its continuity matters. Moreover, how does one find the value function of a given $h^{-1}(.)$? The answer to these questions are given first.
Let $V(.) \in \mathcal{C}({x})$ (respectively, $V(.) \in \mathcal{C}(\Omega)$) denote that function $V(.)$ is continuous at point ${x}$ (respectively, within $\Omega$). 
\begin{Lem} \label{Lemma_Conv_V0} If $h(.)$ is an admissible control policy within $\Omega$, then selecting any $V_{h}^{0}(.)\in \mathcal{C}(\Omega)$ which satisfies $0 \leq V_{h}^{0}(x) \leq U(x,0), \forall x \in \Omega$, the iterations given by
\begin{equation}
		V_{h}^{j+1}(x) = U\big(x,h(x)\big) + V_{h}^{j}\Big(f\big(x,h(x)\big) \Big), \forall x \in \Omega. \label{PI_Value_eq2}
\end{equation}
converges monotonically to the value function of $h(.)$. 
\end{Lem}
\textit{Proof}: It is known that value function $V_h(.)$ is the \textit{fixed point} of iterations indexed by $j$ and given by (\ref{PI_Value_eq2}), \cite{LewisContSystMag}. The reason is Eq. (\ref{PI_Value_eq2}) is equivalent of
\begin{equation}
V_{h}^{j}(x_0)= V_{h}^{0}(x^h_{j}) + \sum_{k=0}^{j-1} {U\big(x^h_{k}, h(x^h_{k})\big)}. \label{V0_Conv_Lem_eq1}
\end{equation}
Comparing (\ref{V0_Conv_Lem_eq1}) with (\ref{ValueFunction_of_h}) and considering $0 \leq V_{h}^{0}(x) \leq U(x,0)$ one has $0 \leq V_{h}^{j}(x) \leq V_h(x)$. Therefore, sequence $\{ V_h^j(x) \}_{j=0}^{\infty}$ is upper bounded by $V_h(x)$. The limit function $V_h^{\infty}(.)$ is equal to $V_h(.)$, since, the admissibility of $h(.)$ leads to $x_{j}^h \to 0$, and hence, $V_{h}^{0}(x_{j}^h) \to 0$ as $j \to \infty$, due to $0 \leq V_{h}^{0}(x) \leq U(x,0)$. Hence, (\ref{V0_Conv_Lem_eq1}) converges to (\ref{ValueFunction_of_h}) as $j \to \infty$. This proves \textit{pointwise} convergence of the sequence to $V_h(.)$.

As for monotonicity of this convergence, not that for any arbitrary positive integers $j_1$ and $j_2$, if $j_1 \leq j_2$, then
\begin{equation}
\begin{split}
V_{h}^{j_1}(& x_0) - V_{h}^{j_2}(x_0)= \\
& V_{h}^{0}(x^h_{j_1}) - V_{h}^{0}(x^h_{j_2}) - \sum_{k=j_1}^{j_2-1} {U\big(x^h_{k}, h(x^h_{k})\big)} \leq 0,  \label{PI_Value_eq3_2}
\end{split}
\end{equation}
since $0\leq V_{h}^{0}(x^h_{j_1}) \leq U\big(x^h_{j_1}, 0\big) \leq U\big(x^h_{j_1}, h(x^h_{j_1})\big)$, and the last term in the foregoing inequality is only one of the non-negative terms in the summation in the right hand side of (\ref{PI_Value_eq3_2}). Therefore, sequence of functions $\{V_{h}^{j}(x)\}_{j=0}^{\infty}$ is pointwise non-decreasing.
\qed 

While the foregoing lemma helps in finding the value function of a given control policy, it does not prove the possible continuity of the result. The reason is, even though $V_h^j(.)$ is continuous for any $j<\infty,$ as a finite sum of continuous functions given in (\ref{V0_Conv_Lem_eq1}), the limit function may not be continuous, as only the \textit{pointwise} convergence is proved. An idea for proof of uniform continuity can be adapted from \cite{Rinehart_VI_TAC}, which is based on \cite{Lincol_RelaxingDynProg}, and hence, requires $V_h\big(f(x,u)\big) \leq \beta U(x,u)$ to hold uniformly for a constant $\beta$. However, 
the foregoing condition restricts the generality of the result. 
Besides, it is not applicable to the assumed case of positive semi-definite $U(.,0)$, because, the value function is positive definite. The following lemma pursues another idea to this end.

\begin{Lem} \label{Lemma_Cont_V0} If $h(.)$ is an admissible control policy within $\Omega$, 
then $V_h(.) \in \mathcal{C}(\Omega)$. 
\end{Lem}
\textit{Proof}: 
The proof is done by contradiction. Assume that $V_h(.)$ is discontinuous at some $y_0\in\Omega$. 
Then
\begin{equation}
\begin{split}
	\exists \epsilon > 0, & \forall \delta > 0, \exists x_0\in\Omega : \\
	& | V_h(x_0) - V_h(y_0)| > \epsilon \mbox{ while } \| x_0 - y_0\| < \delta, \label{Lem_Cont_V0}
\end{split}
\end{equation}
where $\| .\|$ denotes a vector norm and `:' denoted `such that'.
The idea is showing that (\ref{Lem_Cont_V0}) is not possible. To this end, considering recursive relation (\ref{PI_Value_eq2}) initiated with $V_h^0(.)=0$, from (\ref{ValueFunction_of_h}) and (\ref{V0_Conv_Lem_eq1}) one has
\begin{equation}
	V_h(x_0) = V_h^j(x_0) + V_h(x_j^h). \label{Lem_Cont_V1}
\end{equation}
Therefore,
\begin{equation}
	V_h(x_0) - V_h(y_0) = V_h^j(x_0) + V_h(x_j^h) - V_h^j(y_0) - V_h(y_j^h), \label{Lem_Cont_V2}
\end{equation}
which leads to
\begin{equation}
	|V_h(x_0) - V_h(y_0)| \leq |V_h^j(x_0) - V_h^j(y_0)| + V_h(x_j^h) + V_h(y_j^h), \label{Lem_Cont_V3}
\end{equation}
by triangle inequality of absolute values. Inequality (\ref{Lem_Cont_V3}) is the key to the solution, as it will be shown that the right hand side of the inequality can be made arbitrarily small if $x_0$ is close enough to $y_0$ and $j$ is large enough. By $V_h(x) \leq W(x),\forall x$, for some $W(.)\in\mathcal{C}(\Omega), W(0)=0$, per the admissibility of $h(.)$, one has
\begin{equation}
	|V_h(x_0) - V_h(y_0)| \leq |V_h^j(x_0) - V_h^j(y_0)| + W(x_j^h) + W(y_j^h). \label{Lem_Cont_V3_2}
\end{equation}
By admissibility of $h(.)$ one has $y_j^h \to 0$ as $j \to \infty$. Hence, by $W(.)\in\mathcal{C}(\Omega)$ and $W(0)=0$, one has
\begin{equation}
	\begin{split}
\forall y_0 \in  \Omega, \forall \epsilon > 0, \exists & j_1 = j_1(y_0,\epsilon):\\
	& j \geq j_1 \Rightarrow 0 \leq W(y_j^h) < \epsilon/4. \label{Lem_Cont_V4}
\end{split}
\end{equation}
Moreover, by $W(.)\in\mathcal{C}(\Omega)$
\begin{equation}
	\begin{split}
\forall y_j^h \in & \Omega, \forall \epsilon > 0, \exists \delta_1 = \delta_1(y_j^h,\epsilon) :\\
	&\|x_j^h - y_j^h\| < \delta_1 \Rightarrow | W(x_j^h) - W(y_j^h)| < \epsilon/4. \label{Lem_Cont_V5}
\end{split}
\end{equation}
Hence,
\begin{equation}
	\|x_j^h - y_j^h\| < \delta_1 \Rightarrow W(x_j^h) < \epsilon/4 + W(y_j^h). \label{Lem_Cont_V6}
\end{equation}
On the other hand, due to the Lipschitz continuity of the closed loop system, the state trajectory at each finite time, for example $j$, \textit{continuously} depends on the initial conditions, \cite{Khalil}. In other words, $x_j^h$ changes continuously as $x_0$ changes, hence,
\begin{equation}
	\begin{split}
\forall y_0 \in \Omega, \forall \delta_1 > 0, &\exists \delta_2 = \delta_2(y_0,\delta_1,j) :\\
	&\|x_0 - y_0\| < \delta_2 \Rightarrow \| x_j^h - y_j^h\| < \delta_1. \label{Lem_Cont_V7}
\end{split}
\end{equation}
Note that $V_h^j(.)\in\mathcal{C}(\Omega)$, for $j<\infty,$ as mentioned before this lemma. Hence,
\begin{equation}
	\begin{split}
\forall y_0 \in \Omega, &\forall \epsilon > 0, \exists \delta_3 = \delta_3(y_0,\epsilon,j) :\\
	&\|x_0 - y_0\| < \delta_3 \Rightarrow | V_h^j(x_0) - V_h^j(y_0)| < \epsilon/4. \label{Lem_Cont_V8}
\end{split}
\end{equation}
Now we have enough inequalities to contradict (\ref{Lem_Cont_V0}). For any point of discontinuity $y_0$ and $\epsilon$ whose existence is guaranteed by (\ref{Lem_Cont_V0}), find $j_1=j_1(y_0,\epsilon)$ which leads to 
\begin{equation}
	W(y_{j_1}^h) < \epsilon/4, \label{Lem_Cont_V9}
\end{equation}
per (\ref{Lem_Cont_V4}). Then, select $\delta_1 = \delta_1(y_{j_1}^h,\epsilon)$. Per (\ref{Lem_Cont_V6}) and (\ref{Lem_Cont_V9}) one has 
\begin{equation}
	\|x_{j_1}^h - y_{j_1}^h\| < \delta_1 \Rightarrow W(x_{j_1}^h) < \epsilon/4 + W(y_{j_1}^h) < \epsilon/2.
 \label{Lem_Cont_V10}
\end{equation}
Select $\delta_2 = \delta_2(y_0,\delta_1,j_1)$ to have
\begin{equation}
	\|x_0 - y_0\| < \delta_2 \Rightarrow \|x_{j_1}^h - y_{j_1}^h \| < \delta_1,
 \label{Lem_Cont_V11}
\end{equation}
per (\ref{Lem_Cont_V7}). 
Finally, set $\delta_3 = \delta_3(y_0,\epsilon,j_1)$ to have
\begin{equation}
	\|x_0 - y_0\| < \delta_3 \Rightarrow | V_h^{j_1}(x_{j_1}^h) - V_h^{j_1}(y_{j_1}^h)| < \epsilon/4,
 \label{Lem_Cont_V12}
\end{equation}
per (\ref{Lem_Cont_V8}). Select $\delta = \min(\delta_2,\delta_3)$. Using (\ref{Lem_Cont_V9}), (\ref{Lem_Cont_V10}), (\ref{Lem_Cont_V11}), and (\ref{Lem_Cont_V12}) one has
\begin{equation}
\begin{split}
	\|x_0 - y_0\| &< \delta \Rightarrow |V_h(x_0) - V_h(y_0)| \leq \\
	&|V_h^{j_1}(x_0) - V_h^{j_1}(y_0)| + W(x_{j_1}^h) + W(y_{j_1}^h) < \epsilon,
 \label{Lem_Cont_V13}
\end{split}
\end{equation}
which contradicts (\ref{Lem_Cont_V0}). Therefore, $V_h(.)\in\mathcal{C}(\Omega)$.
\qed

Even though we managed to skip the proof of uniform convergence for deriving the desired continuity result, the uniformness of the convergence of (\ref{PI_Value_eq2}), if established, will still be useful. The reason is, the uniform convergence can be used in finding the value function of an admissible control through guaranteeing that there exists a large enough iteration index such that one has
\begin{equation}
|V_h^j(x) - V_{h}^{j+1}(x)| \leq \delta, \forall x \in \Omega,  \label{V0_Unif_Conv_eq1}
\end{equation}
for any selected \textit{constant} tolerance $\delta >0$. This condition can be used for terminating the iterations.
The next lemma proves the desired uniform convergence.

\begin{Lem} \label{Lemma_Cont_W_vs_V} If $h(.)$ is an admissible control policy within compact set $\Omega$, 
then selecting any $V_{h}^{0}(.)\in \mathcal{C}(\Omega)$ which satisfies $0 \leq V_{h}^{0}(x) \leq U(x,0), \forall x \in \Omega$, the iterations given by (\ref{PI_Value_eq2}) converges uniformly in $\Omega$. 
\end{Lem}
\textit{Proof}: Using Dini's uniform convergence theorem (Ref. \cite{Rudin}, Theorem 7.13), the pointwise monotonicity of $\{V_{h}^{j}(x)\}_{j=0}^{\infty}$ (Lemma \ref{Lemma_Conv_V0}), the continuity of the elements of the foregoing sequence, the continuity of the limit function $V_h(.)$ (Lemma \ref{Lemma_Cont_V0}), and the compactness of $\Omega$, lead to the uniform convergence of the iterations. \qed

Now that the concern about continuity of $V_h(.)$ is resolved, the next step is a lemma which proves that the $argmin$ operator will not cause discontinuity in the functions subject to investigation. 
\begin{Lem} \label{Lemma_Cont_W_vs_V} Let $W(x,u) := U(x,u) + V^i\big(f(x,u)\big)$ and $h(x) = argmin_{u\in\mathbb{R}^m} W(x,u)$. If functions $f(.,.), U(.,.),$ and $V^i(.)$ are continuous within $\Omega$, then so is $W\big(.,h(.)\big)$.
\end{Lem}

\textit{Proof}:  
The proof is done by showing that the directional limit of $W\big(.,h(.)\big)$ at any selected point is equal to its evaluation at the point, and hence, it is continuous at that point (motivated by \cite{Heydari_Franklin}).

Let $\bar{x}$ be an arbitrary point in $\Omega$. Set
\begin{equation}
\bar{u}:=h(\bar{x}). \label{Lem3_eq1}
\end{equation}
Select an open set $\alpha \subset \mathbb{R}^n$ such that $\bar{x}$ belongs to the boundary of $\alpha$ and limit
\begin{equation}
\hat{u} := \lim_{x \to \bar{x}, x \in \alpha} h(x), \label{Lem3_eq2}
\end{equation}
exists. If $\bar{u} = \hat{u}$, for every such $\alpha$, then $h(.) \in \mathcal{C}(\bar{x})$. In this case the continuity of $W\big(.,h(.)\big)$ at $\bar{x}$ follows from the continuity of its forming functions, \cite{Rudin}. 

Now assume $\bar{u} \neq \hat{u}$, for some $\alpha$ denoted with $\alpha_0$. From $W(.,\hat{u}) \in \mathcal{C}(\Omega)$ for the given $\hat{u}$, one has
\begin{equation}
W(\bar{x},\hat{u}) = \lim_{x \to \bar{x}, x \in \alpha_0} W(x,\hat{u}), \label{Lem3_eq4}
\end{equation}
If it can be shown that, for every selected $\alpha_0$, one has
\begin{equation}
W(\bar{x},\bar{u}) = W(\bar{x},\hat{u}), \label{Lem3_eq5}
\end{equation}
then the continuity of $W\big(.,h(.)\big)$ at $\bar{x}$ follows, because from (\ref{Lem3_eq4}) and (\ref{Lem3_eq5}) one has
\begin{equation}
W(\bar{x},\bar{u}) = \lim_{x \to \bar{x}} W(x,\hat{u}),
\label{Lem3_eq6}
\end{equation}
and (\ref{Lem3_eq6}) leads to the continuity by definition, \cite{Rudin}. 

\begin{wrapfigure}{r}{0.15\textwidth}
  \vspace{-10pt}
  \begin{center}
		\includegraphics[width=0.2\columnwidth]{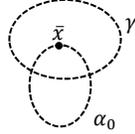}
  \end{center}
  \vspace{-10pt}
		\caption{Schematic of point $\bar{x}$ and open sets $\alpha_0$.} 
		\label{Fig_alpha_gamma_sets}
  \vspace{-10pt}
\end{wrapfigure}

The proof that (\ref{Lem3_eq5}) holds is done by contradiction. 
Assume that for some $\bar{x}$ and some $\alpha_0$ one has
\begin{equation}
W(\bar{x},\bar{u}) > W(\bar{x},\hat{u}). \label{Lem3_eq9}
\end{equation}
Inequality (\ref{Lem3_eq9}) leads to $h(\bar{x}) \neq \bar{u}$. But, this is against (\ref{Lem3_eq1}), hence, (\ref{Lem3_eq9}) cannot hold.
Now, assume 
\begin{equation}
W(\bar{x},\bar{u}) < W(\bar{x},\hat{u}), \label{Lem3_eq7}
\end{equation}
hence there exists some $\epsilon_1 > 0$ such that
\begin{equation}
W(\bar{x},\bar{u}) + \epsilon_1 = W(\bar{x},\hat{u}), \label{Lem3_eq7_1}
\end{equation}
then, due to the continuity of both sides of (\ref{Lem3_eq7_1}) at $\bar{x}$ for the fixed $\bar{u}$ and $\hat{u}$, there exists an open set $\gamma$ containing $\bar{x}$, see Fig. \ref{Fig_alpha_gamma_sets}, and some $\epsilon_2>0$, such that 
\begin{equation}
W(x,\bar{u})  + \epsilon_2 < W(x,\hat{u}),\forall x \in \gamma. \label{Lem3_eq8}
\end{equation}

Given $W\big(x,h(x)\big) \leq  W(x,\bar{u})$, inequality (\ref{Lem3_eq8}) implies that at points which are \textit{close enough} to $\bar{x}$, function $W\big(x,h(x)\big)$ is away from $W(x,\hat{u})$ at least by a margin of $\epsilon_2$. But, this contradicts Eq. (\ref{Lem3_eq2}) which, 
implies that $h(x)$ can be made arbitrarily close to $\hat{u}$ as $x$ gets close to $\bar{x}$ within $\alpha_0$. The reason is, the latter, given the continuity of $W(x,u)$ versus both $x$ and $u$, leads to the conclusion that function $W\big(x,h(x)\big)$ can be made arbitrarily close to $W(x,\hat{u})$ if $x$ approaches $\bar{x}$ from a certain direction. Note that sets $\gamma$ and $\alpha_0$ are not disjoint, as $\bar{x}$ is \emph{within} $\gamma$ and on the \emph{boundary} of $\alpha_0$, as shown in Fig. \ref{Fig_alpha_gamma_sets}. Hence, inequality (\ref{Lem3_eq7}) also cannot hold. 
Therefore, (\ref{Lem3_eq5}) holds and hence, $W(.,h(.)) \in \mathcal{C}(\bar{x})$. Finally, the continuity of the function subject to investigation at any arbitrary $\bar{x} \in \Omega$, leads to the continuity of the function in $\Omega$.
\qed

Now, we have all the required tools to make the following desired conclusion.
\begin{Thm} \label{Thm_Cont_Vi} The value functions at each iteration of stabilizing value iteration are continuous functions, i.e., $V^{i}(.) \in \mathcal{C}(\Omega), \forall i \in \mathbb{N}$.
\end{Thm}
\textit{Proof}: 
The theorem can be proved by induction. Lemma \ref{Lemma_Cont_V0} proves $V^0(.) \in \mathcal{C}(\Omega)$. Assume that $V^i(.) \in \mathcal{C}(\Omega)$. From Lemma \ref{Lemma_Cont_W_vs_V} it follows that $V^{i+1}(.) \in \mathcal{C}(\Omega)$, because, $W(.,h(.)) = V^{i+1}(.)$ where $W(.,h(.))$ is defined in Lemma \ref{Lemma_Cont_W_vs_V}.
\qed

The proof of continuity is of interest for two reasons. 1) to guarantee suitable approximation capability, especially in generalization, i.e., approximating the function at the sample states which were not used during the training stage, \cite{Weierstrass_Theorem, Hornik_NN_Continuity}, and 2) for using the value functions as Lyapunov functions, for proof of stability, as given next.
 
\begin{Thm} \label{Thm_Stab_LyapFun}
Let the compact domain $\mathcal{B}^{i}_r$ for any $r\in\mathbb{R}_+$ be defined as $\mathcal{B}^{i}_r:=\{ x\in\mathbb{R}^n : {V}^{i}(x) \leq r \}$ and let $\bar{r}^i > 0$ be (the largest $r$) such that $\mathcal{B}^{i}_{\bar{r}^i} \subset \Omega$. Then, for every given $i \in \mathbb{N}$, control policy $h^{i}(.)$ resulting from stabilizing value iteration asymptotically stabilizes the system about the origin and $\mathcal{B}^{i}_{\bar{r}^i}$ will be an estimation of the region of attraction for the system.
\end{Thm}

\textit{Proof}: The proof is done by showing that $V^{i}(.)$ is a Lyapunov function for $h^{i}(.)$, for each given $i$. 
Denoting the value function of the initial stabilizing control with $V^0(.)$, it is continuous (Lemma \ref{Lemma_Cont_V0}) and positive definite, by positive semi-definiteness of $U(.,.)$ and Assumption \ref{Assum_InvariantSet}. Note that, there is no $x \neq 0$ with the value function of zero under any control policy. If $V^i(.)$ for some $i$ is positive definite, it directly follows from (\ref{VI_ValueUpdate2}) that $V^{i+1}(.)$ will also be positive definite, because, if $U(x,0)=0$ for some $x \neq 0$, then $f(x,0) \neq x$ by Assumption \ref{Assum_InvariantSet}. Hence, by induction, $V^{i+1}(.)$ is positive definite for every $i \in \mathbb{N}$. Also, by Theorem \ref{Thm_Cont_Vi} it is a continuous function in $\Omega$. 

By (\ref{VI_ValueUpdate2})
\begin{equation}
		V^i\Big(f\big(x,h^i(x)\big)\Big) - V^{i+1}(x) = - U\big(x,h^i(x)\big), \forall x \in \Omega. \label{Thm1_eq1}
\end{equation}
On the other hand, by Lemma \ref{Lemma_NonDecreasing}, inequality (\ref{Lemma1_eq7}) holds for all $i$s. Therefore, replacing $V^{i+1}(x)$ in (\ref{Thm1_eq1}) with $V^{i}(x)$ leads to
\begin{equation}
		V^{i}\Big(f\big(x,h^i(x)\big)\Big) - V^{i}(x) \leq - U\big(x,h^i(x)\big), \forall x \in \Omega. \label{Thm1_eq2}
\end{equation}
Let $S:=\{x \in \mathbb{R}^n : U(x,0) = 0 \}$. The right hand side of (\ref{Thm1_eq2}) can be zero only if $x \in S$. Since, by Assumption \ref{Assum_InvariantSet}, no non-zero state trajectory can stay in $S$, the asymptotic stability of $h^i(.)$ follows from negative semi-definiteness of the difference between the value functions in (\ref{Thm1_eq2}), using LaSalle's invariance theorem, \cite{Khalil}.

Set $\mathcal{B}^{i}_{\bar{r}^i}$ is an EROA \cite{Khalil} for the closed loop system, because, $V^{i}(x_{k+1}) \leq V^i(x_k)$ by (\ref{Thm1_eq2}), hence, $x_k\in\mathcal{B}^{i}_{\bar{r}^i}$ leads to $x_{k+1}\in\mathcal{B}^{i}_{\bar{r}^i}, \forall k \in \mathbb{N}$. 
Finally, since $\mathcal{B}^i_{\bar{r}^i}$ is contained in $\Omega$, it is bounded. Also, the set is closed, because, it is the \textit{inverse image} of a closed set, namely $[0,\bar{r}^i]$ under a continuous function (Theorem \ref{Thm_Cont_Vi}), \cite{Rudin}. Hence, $\mathcal{B}^i_{\bar{r}^i}$ is compact. The origin is an \textit{interior} point of the EROA, because $V^i(0)=0$, $\bar{r}^i>0$, and $V^i(.)\in\mathcal{C}(\Omega)$.
\qed

Theorem \ref{Thm_Stab_LyapFun} proves that each single $h^i(.)$ if \textit{constantly} applied on the system, will have the states converge to the origin. However, in online learning, the control will be subject to adaptation. In other words, if $h^i(.)$ is applied at the current time, control policy $h^{i+1}(.)$ will be applied at the next time-step. It is important to note that even though Theorem \ref{Thm_Stab_LyapFun} proves the asymptotic stability of the \textit{autonomous} system $x_{k+1}=F(x_k):=f\big(x_k,h^i(x_k)\big)$ for every fixed $i$, it does not guarantee the asymptotic stability of the \textit{non-autonomous} system $x_{k+1}=F(x_k,k):=f\big(x_k,h^{k}(x_k)\big)$.
Therefore, it is required to have a separate stability analysis to show that the trajectory formed under the \textit{adapting/evolving} control policy also will converge to zero. An idea for doing that is finding a \textit{single} function, possibly $V^0(.)$, to be a Lyapunov function for \textit{all} the control policies. The proof of the following theorem, however, uses another approach. 

\begin{Thm} \label{Thm_Stabil_EVI_No_Lyap} 
If the system is operated using control policy $h^k(.)$ at time $k$, that is, the control subject to adaptation in the stabilizing value iteration, then, the origin will be asymptotically stable and every trajectory contained in $\Omega$ will converge to the origin.
\end{Thm}
\textit{Proof}: 
Eq. (\ref{VI_ValueUpdate2}) and the monotonicity feature established in Lemma \ref{Lemma_NonDecreasing} lead to
\begin{equation}
\begin{split}
		V^{1}(x_0) = U\big(x_0,h^0(x_0)\big) + V^0\Big(f\big(x_0,h^0(x_0)\big)\Big) \leq \\
		V^0(x_0), \forall x_0 \in \Omega, \label{Thm_Stab_NoLyap_1}
\end{split}
\end{equation}
and similarly
\begin{equation}
\begin{split}
		V^{2}(x) = U\big(x,h^1(x)\big) + V^1\Big(f\big(x,h^1(x)\big)\Big) \leq \\
		V^1(x) \leq V^0(x), \forall x \in \Omega. \label{Thm_Stab_NoLyap_2}
\end{split}
\end{equation}
Let $x_k^* := f\big(x_{k-1}^*,h^{k-1}(x_{k-1}^*)\big)$ for $k \in \mathbb{N} - \{0 \},$ and $x_{0}^*:=x_0$. Evaluating (\ref{Thm_Stab_NoLyap_2}) at $x_1^*$ and replacing the $V^0(x_1^*)$ in the left hand side of the inequality in (\ref{Thm_Stab_NoLyap_1}) with the left hand side of (\ref{Thm_Stab_NoLyap_2}), which is smaller per (\ref{Thm_Stab_NoLyap_2}), one has
\begin{equation}
\begin{split}
		U\big(x_0^*,h^0(x_0^*)\big) + U\big(x_1^*,h^1(x_1^*)\big) + V^1(x_2^*) \leq \\
		V^0(x_0^*), \forall x_0^* \in \Omega. \label{Thm_Stab_NoLyap_3}
\end{split}
\end{equation}
Repeating this process by replacing $V^1(x_2^*)$ in (\ref{Thm_Stab_NoLyap_3}) using
\begin{equation}
\begin{split}
		V^{3}(x_2^*) = U\big(x_2^*,h^2(x_2^*)\big) + V^2(x_3^*) \leq \\
		V^2(x_2^*)  \leq V^1(x_2^*) , \forall x_2^* \in \Omega. \label{Thm_Stab_NoLyap_4}
\end{split}
\end{equation}
leads to
\begin{equation}
\begin{split}
		U\big(x_0^*,h^0(x_0^*)\big) + U\big(x_1^*,h^1(x_1^*)\big) + U\big(x_2^*,h^2(x_2^*)\big) \\
		+ V^2(x_3^*) \leq V^0(x_0), \forall x_0^* \in \Omega. \label{Thm_Stab_NoLyap_5}
\end{split}
\end{equation}
Similarly by repeating this process one has
\begin{equation}
\begin{split}
		\sum_{k=0}^{i-1} U\big(x_k^*,h^k(x_k^*)\big) + V^{i-1}(x_i^*) \leq V^0(x_0), \\
		\forall x_0^* \in \Omega, \forall i \in \mathbb{N}-\{0\}. \label{Thm_Stab_NoLyap_6}
\end{split}
\end{equation}
Since $V^{i-1}(x) \geq 0, \forall x$, the foregoing equations leads
\begin{equation}
		\sum_{k=0}^{i-1} U\big(x_k^*,h^k(x_k^*)\big) \leq V^0(x_0), \forall x_0^* \in \Omega, \forall i \in \mathbb{N}-\{0\}, \label{Thm_Stab_NoLyap_7}
\end{equation}
that is, the sequence of partial sums of the left hand side is upper bounded and because of being non-decreasing,
it converges, as $i\to\infty$, \cite{Rudin}. Therefore, $U\big(x_i^*,h^i(x_i^*)\big) \to 0$ as $i\to\infty$. Considering Assumption \ref{Assum_InvariantSet}, this leads to $x_i^* \to 0$, as long as the entire state trajectory is contained in $\Omega$. 
\qed

It can be seen that Theorem \ref{Thm_Stabil_EVI_No_Lyap} does not provide an EROA. Therefore, the training domain $\Omega$ needs to be selected large enough to guarantee that states, on their way of traveling toward the origin (with not a necessarily straight path) do not exit $\Omega$. However, once the convergence of the value iteration is established, some analytical results regarding the desired EROA will be presented (Theorem \ref{Thm_ROA_Evolving_VI}). 

Besides stability, which is addressed, the convergence of the VI using the stabilizing initial guess also needs to be analyzed. The reason is, none of the cited existing convergence proofs is applicable, as they either require $V^0(x)=0, \forall x,$ \cite{AlTamimi}, or $0 \leq V^0(x) \leq U(x,0), \forall x,$ \cite{Heydari_TCYB}. For example, $0 \leq V^0(x) \leq U(x,0), \forall x,$ does not hold here because $U(x,.)$, which is greater than or equal to $U(x,0)$, is only one of the terms existing in the summation over infinite number of non-negative terms in the definition of $V^0(x)$ as a value function of state $x$. 
As for the convergence result in \cite{Lincol_RelaxingDynProg} whose less restrictive version was presented in \cite{Relaxing_DP_2}, it requires $V^*\big(f(x,u)\big) \leq \beta U(x,u)$ to hold uniformly. This condition is restrictive and not applicable to our analysis, as detailed before Lemma \ref{Lemma_Cont_V0}. 

The main idea for the convergence proof in here is adapted from \cite{Heydari_TCYB}, in which, an analogy was established between the \textit{iterations} of the VI and the \textit{horizon length} of a finite-horizon optimal control problem with a \textit{fixed final time}. Let the cost function for the finite-horizon problem be given by
\begin{equation}
{J}^N=\psi(x_N) + \sum_{k=0}^{N-1} {U(x_k,u_k)}, \label{CostFunction_FinHor}
\end{equation}
where $\psi: \mathbb{R}^n \to \mathbb{R}_+$ is a continuous and positive semi-definite function representing the \textit{terminal cost}. The finite-horizon problem is defined as minimizing ${J}^N$ subject to the dynamics given by (\ref{Dynamics}). Once the final time is fixed, the value function and the control policy become time-dependent \cite{Kirk, Heydari_NN_FinHor}, i.e., they may be denoted with ${V}^{*}(.,.)$ and ${h}^{*}(.,.)$, respectively, where the second argument is the number of remaining time steps, or \textit{time-to-go}. Let the optimal finite-horizon value function given state $x_0$ and time-to-go $\tau \in \mathbb{M}:=\{ 0,1,...,N \}$ be denoted by ${V}^{*}:\mathbb{R}^n \times \mathbb{M} \to \mathbb{R}_+$, where
\begin{equation}
{V}^{*}(x_0,\tau)=\psi(x_{\tau}) + \sum_{k=0}^{\tau-1} {U\big(x^{*,\tau}_k,{h}^{*}(x^{*,\tau}_k,\tau-k)\big)}, \label{ValFunction_FinHor}
\end{equation}
$x^{*,\tau}_k := f\big(x^{*,\tau}_{k-1}, {h}^*(x^{*,\tau}_{k-1}, \tau-(k-1))\big), \forall k$ such that $1 \leq k \leq \tau,$ and $x_0^{*,\tau}:=x_0, \forall \tau$. In other word, the summation is evaluated along the trajectory generated by applying the time varying control policy ${h}^{*}(.,\tau-k)$ at time $k$. Clearly 
\begin{equation}
{V}^{*}(x,0)=\psi(x), \forall x, \label{Bellman_FinHor_eq1}
\end{equation}
and by the Bellman equation for fixed-final-time problems \cite{Kirk, Heydari_NN_FinHor}
\begin{equation}
\begin{split}
{V}^{*}(x,\tau+1)= min_{u\in\mathbb{R}^m} \Big( U(x,u) + {V}^{*}\big(f(x,u),\tau \big) \Big), \\
 \forall x, \forall \tau \in \mathbb{M} - \{N\}, \label{Bellman_FinHor_eq2}
\end{split}
\end{equation}
and
\begin{equation}
\begin{split}
{h}^{*}(x,\tau+1)= & argmin_{u\in\mathbb{R}^m} \Big( U(x,u) + \\
& {V}^{*}\big(f(x,u),\tau \big) \Big),\forall x, \forall \tau \in \mathbb{M} - \{N\}. \label{Bellman_FinHor_eq2_2}
\end{split}
\end{equation}
If $\psi(.)$ in the finite-horizon problem is selected equal to initial guess $V^0(.)$ in the VI, then, comparing Eq. (\ref{Bellman_FinHor_eq2}) with (\ref{VI_ValueUpdate}) it directly follows that
\begin{equation}
{V}^{*}(x,i)= V^i(x), \forall x, \forall i \in \mathbb{M}. \label{FinHor_vs_VI_eq3}
\end{equation}
In other words, the \textit{immature} value function at the $i$th iteration of VI is identical to the \textit{optimal} value function of the fixed-final-time problem of minimizing (\ref{CostFunction_FinHor}) with the final time of $i$, when $\psi(.)=V^0(.)$. Similarly comparing (\ref{Bellman_FinHor_eq2_2}) with (\ref{VI_PolicyUpdate}) it can be seen that
\begin{equation}
{h}^{*}(x,i+1)= h^i(x), \forall x, \forall i \in \mathbb{M}. \label{FinHor_vs_VI_eq3_2}
\end{equation}
Using this idea, it is proved in \cite{Heydari_TCYB} that if $V^0(.)$ is smooth and $0 \leq V^0(x) \leq U(x,0),\forall x,$ then VI converges to $V^*(.)$. The following theorem generalizes the convergence proof to cover the case of $V^0(.)$ being a value function. 

Before proceeding to the theorem, however, it is noteworthy that considering Eq. (\ref{FinHor_vs_VI_eq3}), the stability results given by Theorem \ref{Thm_Stab_LyapFun} resembles a method of stability proof in the model predictive control (MPC) literature \cite{Jadbabaie_MPC_CLF}, in which, a control Lyapunov function is utilized as the terminal cost in the respective finite-horizon problems. Using this idea, it is shown that the closed loop system, under each control calculated for the receding horizons, remains stable. 

\begin{Thm} \label{Theorem_VI_Convergence} The stabilizing value iteration converges to the optimal solution of the infinite-horizon problem within the selected compact domain.
\end{Thm}

\textit{Proof}: Considering the analogy between the iteration of VI and the horizon of a finite-horizon problem given by (\ref{FinHor_vs_VI_eq3}), it can be seen that each $V^i(x_0)$ represents the cost-to-go of applying control sequence $\{h^*(x_0,i), h^*(x_1,i-1), ..., h^*(x_{i-1},1)\}$ for the first $i$ steps, which are the optimal control sequence with respect to cost function (\ref{CostFunction_FinHor}), when $\psi(.)=V^0(.)$ and $N = i$, and applying the stabilizing control sequence $\{ h^{-1}(x_i), h^{-1}(x_{i+1}), ... \}$ for the rest of the (infinite) horizon. The reason for this conclusion is the fact that $V^0(x_i)$, which is used as the terminal cost in the fixed-final-time problem itself represents the cost-to-go of applying admissible control policy $h^{-1}(.)$ for infinite number of times, starting from $x_i$, per Eq. (\ref{ValueFunction_of_h}).

On the other hand, the non-increasing (cf. Lemma \ref{Lemma_NonDecreasing}) and non-negative (cf. proof of Theorem \ref{Thm_Stab_LyapFun}) nature of value functions under VI, and hence, of the finite-horizon value function ${V}^*(.,i)$ lead to the convergence of the sequence of value functions to a finite limit function, denoted with $V^{\infty}(.)=V^*(.,\infty)$.
Because, every non-increasing and lower bounded sequence converges, \cite{Rudin}.
Therefore, one has 
\begin{equation}
\lim_{i \to \infty} x_i \to 0, \label{Thm2_eq1}
\end{equation}
using control sequence $\{h^*(x_0,i), h^*(x_1,i-1), ..., h^*(x_{i-1},1)\}$. Otherwise, $V^{\infty}(.)$ becomes unbounded.  
This can also be concluded by noting that the tail of a convergent series can be made arbitrarily small (cf. p. 59 \cite{Rudin}).
Note that per Assumption \ref{Assum_InvariantSet} the state trajectory cannot hide in the invariant set of $f(.,0)$ with zero utility function, to lead to a finite cost-to-go without convergence to the origin.

Due to the continuity and positive semi-definiteness of $V^0(.)$, one has $V^0(.) \to 0$ as $x \to 0$. 
Therefore, by Eq. (\ref{Thm2_eq1}) one has
\begin{equation}
\lim_{i \to \infty} V^0(x_i) \to 0, \label{Thm2_eq2}
\end{equation}
in calculation of the cost-to-go ${V}^*(.,i)$. Comparing finite-horizon cost function (\ref{CostFunction_FinHor}) with infinite-horizon cost function (\ref{CostFunction}) and considering (\ref{Thm2_eq2}), one has
\begin{equation}
V^*(x)=V^{\infty}(x), \forall x \in \Omega. \label{Thm2_eq3}
\end{equation}
Otherwise, the smaller value among $V^*(x)$ and $V^{\infty}(x)$ will be both the optimal value function (evaluated at $x$) for the infinite-horizon problem and the greatest lower bound of the sequence of value function of the fixed-final-time problems resulting from $N = 0, 1, 2, ...$.
\qed

Comparing the results given by Theorems \ref{Thm_Stab_LyapFun} and \ref{Theorem_VI_Convergence} with the existing literature, the closest one is \cite{Liu_Stable_VI}, in which a VI algorithm, called $\theta$-ADP, was introduced. The point that $\theta$-ADP requires to be initiated from a control Lyapunov function (CLF) of the respective system in order for the control under iterations to be stabilizing corresponds to the required initial admissible guess for VI in this study. The reason is, if the CLF is known, an asymptotically stable control can be derived directly from the function, e.g., using Sontag's formula, \cite{Sontag}.
An asymptotically stable control law, however, is not required to lead to a finite cost-to-go. Hence, at the first glance, the condition in $\theta$-ADP seems to be less restrictive compared to the condition of using admissible initial control in this study. However, once the second requirement of $\theta$-ADP, that is the existence and utilization of a scale factor $\theta$ using which 
the CLF function evaluated along the state trajectory decays faster than a value function, 
is taken into account, the control resulting from the scaled CLF will lead to a finite cost-to-go. 
Therefore, the results look similar in regards to the initial guess. However, the simplicity of the proofs especially for the convergence proof, including the intermediate steps for firm conclusions (e.g., continuity analysis), admitting a positive semi-definite running cost as opposed to the positive definite one in that work, and establishing an EROA are the main differences of the mentioned theorems compared with \cite{Liu_Stable_VI}. 

Besides addressing the convergence concern in VI, the foregoing theorem provides an idea for establishing an EROA for the system operated using evolving control policies, as presented next. 

\begin{Thm} \label{Thm_ROA_Evolving_VI} 
Let $\mathcal{B}^{i}_r:=\{ x\in\mathbb{R}^n : {V}^{i}(x) \leq r \}$ and $\mathcal{B}^*_r:=\{ x\in\mathbb{R}^n : {V}^*(x) \leq r \}$ for any $r\in\mathbb{R}_+$. Also, let the system be operated using control policy $h^k(.)$ at time $k$, that is, the control subject to adaptation in the stabilizing value iterations. 
If $\mathcal{B}^{*}_{r} \subset \Omega$ for an $r>0$ then $\mathcal{B}^{0}_{r}$ is an estimation of the region of attraction of the closed loop system.
\end{Thm}
\textit{Proof}: As the first step we show that for any given $r$ one has
\begin{equation}
x_k \in \mathcal{B}^{k}_{r} \Rightarrow x_{k+1}=f\big(x_k,h^k(x_k)\big) \in \mathcal{B}^{k+1}_{r}, \forall k \in \mathbb{N}, \forall r \in\mathbb{R}_+. \label{eq_Thm_ROA_Evolving_VI_1}
\end{equation}
By inequality (\ref{Thm1_eq2}) one has $V^k(x_{k+1}) \leq V^k(x_{k})$. Therefore, 
\begin{equation}
x_k \in \mathcal{B}^{k}_{r} \Rightarrow x_{k+1}=f\big(x_k,h^k(x_k)\big) \in \mathcal{B}^{k}_{r}, \forall k \in \mathbb{N}, \forall r \in\mathbb{R}_+. \label{eq_Thm_ROA_Evolving_VI_2}
\end{equation}
By definition of $\mathcal{B}^{k}_r$ if $V^{k+1}(x) \leq V^{k}(x), \forall x$, which follows from Lemma \ref{Lemma_NonDecreasing}, then $\mathcal{B}^{k}_r \subset \mathcal{B}^{k+1}_r$. Therefore, 
\begin{equation}
x_{k+1} \in \mathcal{B}^{k}_{r} \Rightarrow x_{k+1} \in \mathcal{B}^{k+1}_{r}, \forall k \in \mathbb{N}, \forall r \in\mathbb{R}_+. \label{eq_Thm_ROA_Evolving_VI_3}
\end{equation} 
Finally (\ref{eq_Thm_ROA_Evolving_VI_2}) and (\ref{eq_Thm_ROA_Evolving_VI_3}) lead to (\ref{eq_Thm_ROA_Evolving_VI_1}). Now that (\ref{eq_Thm_ROA_Evolving_VI_1}) is proved, one may use mathematical induction to see
\begin{equation}
x_0 \in \mathcal{B}^{0}_{r} \Rightarrow x_{k} \in \mathcal{B}^{k}_{r}, \forall k \in \mathbb{N}, \forall r \in\mathbb{R}_+. \label{eq_Thm_ROA_Evolving_VI_3_1}
\end{equation}

The next step is noticing that $V^{*}(x) \leq V^{i}(x), \forall x$, which follows from the monotonicity of $\{ V^i(.) \}_{i=0}^{\infty}$ and its convergence to $V^*(.)$, per Theorem \ref{Theorem_VI_Convergence}. The foregoing inequality leads to $\mathcal{B}^{k}_r \subset \mathcal{B}^{*}_r, \forall k$, by definition of $\mathcal{B}^{k}_r$ and $\mathcal{B}^{*}_r$. Therefore, (\ref{eq_Thm_ROA_Evolving_VI_3_1}) leads to 
\begin{equation}
x_0 \in \mathcal{B}^{0}_{r} \Rightarrow x_{k} \in \mathcal{B}^{*}_{r}, \forall k \in \mathbb{N}, \forall r \in\mathbb{R}_+. \label{eq_Thm_ROA_Evolving_VI_4}
\end{equation}
The result given by (\ref{eq_Thm_ROA_Evolving_VI_4}) proves the theorem, because, if $r$ is such that $\mathcal{B}^{*}_{r} \subset \Omega$ then any trajectory initiated within $\mathcal{B}^{0}_{r}$ will remain inside $\Omega$, and hence, by Theorem \ref{Thm_Stabil_EVI_No_Lyap} will converge to the origin.
\qed

\section{Analysis of Stabilizing Approximate Value Iteration}  \label{AVI}

The problem with the exact VI is the issue that \textit{exact} reconstruction of the right hand side of Eq. (\ref{VI_ValueUpdate}) is not generally possible except for very simple problems. In general, parametric function approximators are used for this purpose, which hence, give rise to function approximation errors. 
When the approximation errors are considered, Eq. (\ref{VI_ValueUpdate}) reads  
\begin{equation}
\begin{split}
		\hat{V}^{i+1}(x) = min_{u\in\mathbb{R}^m} \Big( U(x,u) + \hat{V}^i\big(f(x,u)\big)\Big) + \\
		\epsilon^i(x), \forall x \in \Omega, \label{AVI_1}
\end{split}
\end{equation}
where the \textit{approximate value function} at the $i$th iteration is denoted with $\hat{V}^i(.)$ and the approximation error at this iteration is denoted with $\epsilon^i(.)$. Note that the value function in the right hand side of Eq. (\ref{AVI_1}) is also an approximate quantity, generated from the previous iteration. 
When $\epsilon^i(.) \neq 0$, the convergence of the approximate VI (AVI) does not follow from Theorem \ref{Theorem_VI_Convergence}. This convergence/boundedness 
is investigated in this section. 

Before proceeding to the convergence/boundedness analysis it is worth mentioning that one typically trains a control approximator (actor) to approximate the solution to the minimization problem given by (\ref{Bellman_eq2}) based on the value function resulting from VI. The control approximator, will hence, lead to \textit{another} approximation error term in the process, regardless of whether the value function reconstruction is exact or approximate. However, the effect of the actor's approximation error can be removed from both the convergence analysis of the AVI and the stability analysis of the system during value iterations, 
as the control will be directly calculated from the minimization of the right hand side of Eq. (\ref{AVI_1}) in \textit{online} and \textit{adaptive} optimal control and applied on the system. In other words, even though the actor will be updated simultaneously along with the critic in online learning, the critic training and the system's operation are independent of the actor's approximation accuracy. Once the learning is concluded (and if it is concluded), the operation of the system could be based on the control resulting from the trained actor, hence, the actor's approximation error can affect the stability of the system at that stage. The stability analysis \textit{after} conclusion of AVI is beyond the scope of this study and deserves to be investigated separately, as the focus in this work is analyzing the convergence/boundedness and stability \textit{during} the online learning process through AVI.

Considering the above comment and denoting the minimizer of the right hand side of Eq. (\ref{AVI_1}) by $\hat{h}^i(.)$, one has
\begin{equation}
		\hat{h}^{i}(x) = argmin_{u\in\mathbb{R}^m} \Big( U(x,u) + \hat{V}^i\big(f(x,u)\big)\Big), \forall x \in \Omega, \label{AVI_2}
\end{equation}
therefore, Eq. (\ref{AVI_1}) can be written as
\begin{equation}
		\hat{V}^{i+1}(x) = U\big(x,\hat{h}^i(x)\big) + \hat{V}^i\Big(f\big(x,\hat{h}^i(x)\big)\Big) + \epsilon^i(x), \forall x \in \Omega. \label{AVI_3}
\end{equation}

Assuming an upper bound for the approximation error $\epsilon^i(x)$ 
the results given by Theorem \ref{Thm_Boundedness} can be obtained, 
in terms of boundedness of sequence $\{ \hat{V}^i(x) \}_{i=0}^\infty$ resulting from the AVI and its relation versus the optimal value function. This boundedness will later be used for stability analysis.  
 
For our AVI analyses, i.e., for the rest of this study, it is assumed that state penalizing function $Q(.)$ in $U(x,u)=Q(x)+u^TRu$ only vanishes at the origin. In other words, instead of the positive semi-definiteness of $U(.,0)$, it is assumed that $U(.,0)$ is positive definite hereafter. The reason for this modification is the point that $U(x,0)$ is going to be used to put an upper bound on $|\epsilon^i(x)|$, which makes sense only if it does not vanish at any non-zero $x$.     

\begin{Thm} \label{Thm_Boundedness} Let $| \epsilon^i(x)| \leq c U(x,0), \forall x \in \Omega, \forall i \in \mathbb{N},$ for some $c\in [0,1)$.
If the approximate value iteration is initiated using some $\hat{V}^0(x)$ which satisfies $\underline{V}^0(x) \leq \hat{V}^0(x) \leq \overline{V}^0(x), \forall x \in \Omega$ where $\underline{V}^0(x)$ and $\overline{V}^0(x)$ are, respectively, the initial guesses for the exact value iterations corresponding to cost functions
\begin{equation}
\underline{J}=\sum_{k=0}^\infty \Big( U(x_k,u_k) - cU(x_k,0) \Big), \label{Cost_Lower_1}
\end{equation} 
and
\begin{equation}
\overline{J}=\sum_{k=0}^\infty \Big( U(x_k,u_k) + cU(x_k,0) \Big), \label{Cost_Upper_1}
\end{equation}
subject to dynamics (\ref{Dynamics}), then, the result of the approximate value iteration at the $i$th iteration is bounded from below by the result of the exact value iteration corresponding to cost function (\ref{Cost_Lower_1}) and from above by the result of the exact value iteration corresponding to cost function (\ref{Cost_Upper_1}). 
\end{Thm}

\textit{Proof}:
Let $\{ \overline{V}^i(x) \}_{i=0}^\infty$ and $\{ \underline{V}^i(x) \}_{i=0}^\infty$ where $\overline{V}^i:\mathbb{R}^n\to\mathbb{R_+}$ and $\underline{V}^i:\mathbb{R}^n\to\mathbb{R_+}$, be defined as sequences of functions initiated from some $\overline{V}^0(.)$ and $\underline{V}^0(.)$ and generated by  
\begin{equation}
\begin{split}
		\underline{V}^{i+1}(x) = min_{u\in\mathbb{R}^m} \Big( U&(x,u) -cU(x,0) \\
		&+\underline{V}^i\big(f(x,u)\big)\Big), \forall x \in \Omega, \label{V_Lower_1}
\end{split}
\end{equation}
\begin{equation}
\begin{split}
		\overline{V}^{i+1}(x) = min_{u\in\mathbb{R}^m} \Big( U&(x,u) +cU(x,0) \\
		&+\overline{V}^i\big(f(x,u)\big)\Big), \forall x \in \Omega. \label{V_Upper_1}
\end{split}
\end{equation}
Considering recursive relations (\ref{V_Upper_1}) and (\ref{V_Lower_1}) it is seen that $\overline{V}^i(.)$ and $\underline{V}^i(.)$ are, respectively, the value functions at the $i$the iteration of \textit{exact VI} for cost functions (\ref{Cost_Lower_1}) and (\ref{Cost_Upper_1}). Considering this point, the lemma can be proved using mathematical induction. Initially $\hat{V}^0(x) \leq \overline{V}^0(x), \forall x \in \Omega$ by assumption. Let $\hat{V}^i(x) \leq \overline{V}^i(x), \forall x \in \Omega$ hold for some $i$. Comparing Eq. (\ref{V_Upper_1}) with Eq. (\ref{AVI_1}) it follows that $\hat{V}^{i+1}(x) \leq \overline{V}^{i+1}(x)$, since $ \epsilon^i(x) \leq c U(x,0)$ and $\hat{V}^i(x) \leq \overline{V}^i(x), \forall x$. Therefore, one has $\hat{V}^i(x) \leq \overline{V}^i(x), \forall i\in \mathbb{N}, \forall x$. The proof of $\underline{V}^i(x) \leq \hat{V}^i(x), \forall i\in \mathbb{N}$ is similar by induction, through comparing Eq. (\ref{V_Lower_1}) with Eq. (\ref{AVI_1}) and noting that $- c U(x,0) \leq \epsilon^i(x), \forall x, \forall i$. 
\qed

The result given by the foregoing theorem resembles the idea of Relaxed Dynamic Programming presented in \cite{Lincol_RelaxingDynProg}. However, the idea, the proof, and the applications of the boundedness result, presented in the rest of this study, are different. 

The exact VIs given by (\ref{V_Lower_1}) and (\ref{V_Upper_1}) converge, based on Theorem \ref{Theorem_VI_Convergence}, when initiated using value functions (defined based on the respective cost function) of some admissible controls. Therefore, considering Theorem \ref{Thm_Boundedness}, the boundedness of the AVI results for all iterations follows, assuming the boundedness of the approximation errors by $c U(x,0)$ for some $c\in [0,1)$.

The actual convergence as well as stability of the system operated under AVI are much more challenging compared to the respective analyses in exact VI, since the presence of the approximation error cancels the monotonicity feature presented in Lemma \ref{Lemma_NonDecreasing}. Note that the monotonicity was the backbone of both the stability and the convergence results given in Theorems \ref{Thm_Stab_LyapFun}, \ref{Thm_Stabil_EVI_No_Lyap}, and \ref{Theorem_VI_Convergence}. As long as the boundedness of the functions during AVI is guaranteed in a neighborhood  of the optimal value function (Theorem \ref{Thm_Boundedness}) where the neighborhood shrinks if the approximation error decreases, the actual convergence of the iteration may not be of a critical importance in implementing the AVI. But, the stability of the system operated under the AVI is definitely critical. The following lemma develops a `semi-monotonicity' of the stabilizing AVI to be used later for deriving some stability results. Note that, following Definition \ref{StabilizingVI_Definition}, \textit{stabilizing AVI} is defined as the AVI which is initiated using the \textit{approximate} value function of an admissible control policy, with an approximation error denoted with $\epsilon^{-1}(.)$.
In other words instead of the exact value function $V^0(.)$ given by (\ref{VI_Value_eq1}) one initiates the iterations using the approximate value function $\hat{V}^0(.)$ which satisfies
\begin{equation}
\begin{split}
		\hat{V}^{0}(x) = U(x,h^{-1}(x)) + \hat{V}^0\Big(f\big(x,h^{-1}(x)\big)\Big) \\
		+ \epsilon^{-1}(x), \forall x \in \Omega. \label{AVI_V0_eq}
\end{split}
\end{equation}

\begin{Lem} \label{Lemma_AVI_SemiMonotonicity} Let $\breve{V}^{i}(x_0) := \sum_{k=0}^{i} U(\hat{x}_k^{*,i},0), \forall x_0 \in \Omega, \forall i\in\mathbb{N}$, where $\hat{x}_k^{*,i}:=f\big(\hat{x}_{k-1}^{*,i}, \hat{h}^{i-k}(\hat{x}_{k-1}^{*,i}) \big)$ and $\hat{x}_0^{*,i}:=x_0, \forall i, \forall k \in \mathbb{N}-\{0\}$. If the stabilizing approximate value iteration scheme is conducted using a function approximator that satisfies $| \epsilon^i(x)| \leq c U(x,0), \forall i \in \mathbb{N} \cup \{-1\}, \forall x,$ for some $c\in [0,1)$, then,
\begin{equation}
		\hat{V}^{i+1}(x) \leq \hat{V}^{i}(x) + 2c \breve{V}^i(x), \forall x \in \Omega. \label{eq_Lem_AVI_SemiMonot_1}
\end{equation}
\end{Lem}
\textit{Proof}: Initially note that $\hat{x}_k^{*,i}$ is the $k$th state vector on the state trajectory initiated from $x_0$ and propagated by applying control policy $\hat{h}^{(i-1)-\bar{k}}(.)$ at time $\bar{k}, 0 \leq \bar{k} \leq i$. The first iteration of AVI leads to
\begin{equation}
		\hat{V}^{1}(x) = min_{u\in\mathbb{R}^m} \Big( U(x,u) + \hat{V}^0\big(f(x,u)\big)\Big) + \epsilon^0(x), \forall x \in \Omega. \label{eq_Lem_AVI_SemiMonot_2}
\end{equation}
One has $\hat{V}^1(x) - \epsilon^0(x) \leq \hat{V}^0(x) -  \epsilon^{-1}(x), \forall x$, because,  per (\ref{eq_Lem_AVI_SemiMonot_2}), $\hat{V}^1(x)$ is resulted from a minimization, as opposed to using a given policy $h^{-1}(.)$ in (\ref{AVI_V0_eq}). The foregoing inequality along with $-c U(x,0) \leq -\epsilon^i(x) \leq c U(x,0), \forall i,$ lead to 
\begin{equation}
		\hat{V}^1(x) \leq \hat{V}^0(x) + 2c U(x,0), \forall x, \label{eq_Lem_AVI_SemiMonot_3}
\end{equation}
which confirms that inequality (\ref{eq_Lem_AVI_SemiMonot_1}) holds for $i=0$. Now, assume that
\begin{equation}
		\hat{V}^{i}(x) \leq \hat{V}^{i-1}(x) + 2c \breve{V}^{i-1}(x), \forall x \in \Omega, \label{eq_Lem_AVI_SemiMonot_4}
\end{equation}
if this assumption leads to (\ref{eq_Lem_AVI_SemiMonot_1}) the proof will be complete by induction. Let
\begin{equation}
		\hat{\mathcal{V}}(x) := U(x,\hat{h}^{i-1}(x)) + \hat{V}^{i}\Big(f\big(x,\hat{h}^{i-1}(x)\big)\Big) + \epsilon^{i}(x), \forall x \in \Omega. \label{eq_Lem_AVI_SemiMonot_5}
\end{equation}
Since the minimizer of the right hand side of (\ref{AVI_1}) is $\hat{h}^i(.)$ and not $\hat{h}^{i-1}(.)$, one has
\begin{equation}
		\hat{V}^{i+1}(x) \leq \hat{\mathcal{V}}(x), \forall x \in \Omega. \label{eq_Lem_AVI_SemiMonot_6}
\end{equation}
On the other hand, 
by definition of $\hat{V}^{i}(x_0)$, that is,
\begin{equation}
\begin{split}
		\hat{V}^{i}(x) = U(x,\hat{h}^{i-1}(x)) + \hat{V}^{i-1}\Big(f\big(x,\hat{h}^{i-1}(x)\big)\Big) \\
		+ \epsilon^{i-1}(x), \forall x \in \Omega, \label{eq_Lem_AVI_SemiMonot_7}
\end{split}
\end{equation}
and comparing it with (\ref{eq_Lem_AVI_SemiMonot_5}) and considering (\ref{eq_Lem_AVI_SemiMonot_4}) one has
\begin{equation}
\begin{split}
		\hat{\mathcal{V}}(x) - & \epsilon^{i}(x) \leq \hat{V}^{i}(x) - \epsilon^{i-1}(x) \\
		& + 2c \breve{V}^{i-1}\Big(f\big(x,\hat{h}^{i-1}(x)\big)\Big), \forall x \in \Omega, \label{eq_Lem_AVI_SemiMonot_7}
\end{split}
\end{equation}
which because of $-c U(x,0) \leq -\epsilon^i(x) \leq c U(x,0), \forall i,$ leads to
\begin{equation}
		\hat{\mathcal{V}}(x) \leq \hat{V}^{i}(x) +2cU(x,0) + 2c \breve{V}^{i-1}\Big(f\big(x,\hat{h}^{i-1}(x)\big)\Big), \forall x \in \Omega, \label{eq_Lem_AVI_SemiMonot_8}
\end{equation}
and along with (\ref{eq_Lem_AVI_SemiMonot_6}) leads to 
\begin{equation}
\begin{split}
		\hat{V}^{i+1}(x) \leq & \hat{V}^{i}(x) +2cU(x,0) \\
		& +2c \breve{V}^{i-1}\Big(f\big(x,\hat{h}^{i-1}(x)\big)\Big), \forall x \in \Omega. \label{eq_Lem_AVI_SemiMonot_9}
\end{split}
\end{equation}
The next step is showing that 
\begin{equation}
U(x_0,0) + \breve{V}^{i-1}\Big(f\big(x_0,\hat{h}^{i-1}(x_0)\big)\Big) = \breve{V}^{i}(x_0), \forall x_0 \in \Omega. \label{eq_Lem_AVI_SemiMonot_10}
\end{equation}
Note that, $\breve{V}^{i}(x_0)$ is the result of evaluating a finite sum of $U(x_k,0)$'s along a `trajectory' initiated from $x_0$. So, in order to show that (\ref{eq_Lem_AVI_SemiMonot_10}) holds, it suffices to show that the summations in both sides of (\ref{eq_Lem_AVI_SemiMonot_10}), which each has $i+1$ elements, are along the same trajectory.
 
The first summand in $\breve{V}^{i}(x_0)$ is $U(x_0,0)$ which is matched by the same term existing in the left hand side of (\ref{eq_Lem_AVI_SemiMonot_10}). The second summand of $\breve{V}^{i}(x_0)$ is $U(.,0)$ evaluated at $\hat{x}_1^{*,i} = f\big(\hat{x}_{0}^{*,i}, \hat{h}^{i-1}(\hat{x}_{0}^{*,i}) \big)$. The first summand of $\breve{V}^{i-1}\big(f(x_0,\hat{h}^{i-1}(x_0))\big)$ is $U(.,0)$ evaluated at $x_1:=f\big(x_0,\hat{h}^{i-1}(x_0)\big)$. Since $\hat{x}_{0}^{*,i} = x_0$, one has $\hat{x}_{1}^{*,i} = x_1$, hence, the second summand of $\breve{V}^{i}(x_0)$ also will be matched by a term in the left hand side of (\ref{eq_Lem_AVI_SemiMonot_10}). Similarly, the third summand of $\breve{V}^{i}(x_0)$ is $U(.,0)$ evaluated at $\hat{x}_2^{*,i} = f\big(\hat{x}_{1}^{*,i}, \hat{h}^{i-2}(\hat{x}_{1}^{*,i}) \big)$. The second summand of $\breve{V}^{i-1}\big(f(x_0,\hat{h}^{i-1}(x_0))\big)$ is $U(.,0)$ evaluated at $x_2:=f\big(x_1,\hat{h}^{i-2}(x_1)\big)$, by definition. Since $\hat{x}_1^{*,i}=x_1$ one has $\hat{x}_2^{*,i}=x_2$. Repeating this argument it is seen that the trajectories are identical and hence, (\ref{eq_Lem_AVI_SemiMonot_10}) holds, which along with (\ref{eq_Lem_AVI_SemiMonot_9}) proves (\ref{eq_Lem_AVI_SemiMonot_1}).
\qed
  
\begin{Thm} \label{Theorem_Stability_AVI} 
Let the stabilizing approximate value iteration be conducted using a continuous function approximator with the bounded approximation error $| \epsilon^i(x)| \leq c U(x,0), \forall i \in \mathbb{N} \cup \{-1\},\forall x \in \Omega,$ for some $c\in [0,1)$. Moreover, denoting the value function of $h^{-1}(.)$ with $V^0(.)$, let $\gamma \in \mathbb{R}_+$ be such that $V^0(x)
 \leq \gamma U(x,0), \forall x \in \Omega$. 
Then, for every given $i \in \mathbb{N}$, control policy $\hat{h}^{i}(.)$ asymptotically stabilizes the system about the origin if $c$ is such that 
\begin{equation}
		0 \leq c < 1 + 2\gamma - \sqrt{4\gamma^2 + 4 \gamma}. \label{AVI_Stabil_Thm_eq1}
\end{equation}
Moreover, let the compact domain $\hat{\mathcal{B}}^{i}_r$ for any $r\in\mathbb{R}_+$ be defined as $\hat{\mathcal{B}}^{i}_r:=\{ x\in\mathbb{R}^n : \hat{V}^{i}(x) \leq r \}$ and let $\bar{r}^i > 0$ be (the largest $r$) such that $\hat{\mathcal{B}}^{i}_{\bar{r}^i} \subset \Omega$. Then, $\hat{\mathcal{B}}^{i}_{\bar{r}^i}$ will be an estimation of the region of attraction for the system.
\end{Thm}
\textit{Proof}: The idea, similar to Theorem \ref{Thm_Stab_LyapFun}, is using $\hat{V}^i(.)$ as a Lyapunov function to prove the claim. The lower and upper boundedness of the function, established in Theorem \ref{Thm_Boundedness}, guarantees the positive definiteness of the function and the continuity of the parametric function approximator guarantees its continuity. The objective is showing negative definiteness of $\Delta \hat{V}^i(x):= \hat{V}^i\big(f(x,\hat{h}^i(x))\big) - \hat{V}^{i}(x)$. 
 By Eq. (\ref{AVI_3})
\begin{equation}
		\hat{V}^i\Big(f\big(x,\hat{h}^i(x)\big)\Big) - \hat{V}^{i+1}(x) = - U\big(x,\hat{h}^i(x)\big) - \epsilon^i(x), \forall x \in \Omega. \label{AVI_Stabil_Thm_eq2}
\end{equation}
Lemma \ref{Lemma_AVI_SemiMonotonicity} and inequality (\ref{eq_Lem_AVI_SemiMonot_1}) may be used in the foregoing equation to replace $\hat{V}^{i+1}(x)$ with $\hat{V}^{i}(x)$ in its left hand side. Before that, considering the similarity between the exact and approximate VI, from Eqs. (\ref{ValFunction_FinHor}), (\ref{FinHor_vs_VI_eq3}), (\ref{FinHor_vs_VI_eq3_2}), and (\ref{AVI_1}), it is straight forward to see
\begin{equation}
\begin{split}
		\hat{V}^i(x_0) = \hat{V}^0(\hat{x}^{*,i}_i) &+ \\
		\sum_{k=0}^{i-1} \Big(U\big(\hat{x}_k^{*,i},&\hat{h}^{(i-1)-k}(\hat{x}_k^{*,i})\big) + \epsilon^{(i-1)-k}(\hat{x}_k^{*,i}) \Big), \\
		& \forall x_0 \in \Omega, \forall i \in \mathbb{N}-\{0\}, \label{AVI_Stabil_Thm_eq3}
\end{split}
\end{equation}
where $\hat{x}_k^{*,i}:=f\big(\hat{x}_{k-1}^{*,i}, \hat{h}^{i-k}(\hat{x}_{k-1}^{*,i}) \big)$ and $\hat{x}_0^{*,i}:=x_0, \forall i, \forall k \in \mathbb{N}-\{0\}.$ The point in concluding (\ref{AVI_Stabil_Thm_eq3}) from the exact VI counterpart, given by (\ref{ValFunction_FinHor}) considering (\ref{FinHor_vs_VI_eq3}) and (\ref{FinHor_vs_VI_eq3_2}), is the fact that at time $k$ when the control policy is $\hat{h}^{(i-1)-k}(.)$, the approximation error introduced in the summation will be $\epsilon^{(i-1)-k}(.)$, i.e., the superscripts will the same per (\ref{AVI_3}) and both functions will be evaluated at the current state $\hat{x}_k^{*,i}$.

Considering $(1-c) U(x,0) \leq U(x,u) + \epsilon^i(x),\forall x, \forall u, \forall i,$ and comparing (\ref{AVI_Stabil_Thm_eq3}) with $\breve{V}(x_0)$ defined in Lemma \ref{Lemma_AVI_SemiMonotonicity} one has
\begin{equation}
		(1-c)\breve{V}^{i}(x) \leq \hat{V}^i(x), \forall x \in \Omega. \label{AVI_Stabil_Thm_eq4}
\end{equation}
Moreover, by Theorem \ref{Thm_Boundedness} one has $\hat{V}^i(x) \leq \overline{V}^i(x), \forall x,$ if $\overline{V}^i(.)$ is generated using the value function of $h^{-1}(.)$ as the initial guess. Moreover, $\overline{V}^i(x) \leq \overline{V}^0(x), \forall x,$ by Lemma \ref{Lemma_NonDecreasing}. 
Therefore,
\begin{equation}
		\breve{V}^{i}(x) \leq \frac{1}{1-c} \overline{V}^0(x), \forall x \in \Omega, \forall i \in \mathbb{N}. \label{AVI_Stabil_Thm_eq5}
\end{equation}
The interesting point about inequality (\ref{AVI_Stabil_Thm_eq5}) is showing the boundedness of the left hand side. This boundedness along with the results from Lemma \ref{Lemma_AVI_SemiMonotonicity} lead to
\begin{equation}
		\hat{V}^{i+1}(x) \leq \hat{V}^i(x) + \frac{2c}{1-c} \overline{V}^0(x), \forall x \in \Omega, \forall i \in \mathbb{N}. \label{AVI_Stabil_Thm_eq6}
\end{equation}
Utilizing (\ref{AVI_Stabil_Thm_eq6}) in (\ref{AVI_Stabil_Thm_eq2}) leads to
\begin{equation}
\begin{split}
		\hat{V}^i\Big(f\big(x,\hat{h}^i(x)\big)\Big) - \hat{V}^{i}&(x) \leq \\
		 - U\big(x,\hat{h}^i(x)\big)  -& \epsilon^i(x) + \frac{2c}{1-c} \overline{V}^0(x), \forall x \in \Omega. \label{AVI_Stabil_Thm_eq7}
\end{split}
\end{equation}
in order to have $\Delta \hat{V}^i(x) < 0$ one needs $2c/(1-c) \overline{V}^0(x) < U\big(x,\hat{h}^i(x)\big) + \epsilon^i(x), \forall x$, which holds if
\begin{equation}
\begin{split}
		\frac{2c}{1-c} \overline{V}^0(x) & < (1-c) U(x,0) \Leftrightarrow \\
		&\frac{2c}{(1-c)^2}  < \frac{U(x,0)}{\overline{V}^0(x)}, \forall x \in \Omega, \label{AVI_Stabil_Thm_eq8}
\end{split}
\end{equation}
Note that $\overline{V}^0(x) \leq 2 V^0(x), \forall x,$ by definition of $\overline{V}^0(x)$ which is the value function of $h^{-1}(.)$ with the utility of $U(x_k,u_k) + cU(x_k,0)$, while, $V^0(x)$ is the value function of the same control policy with the utility of $U(x_k,u_k)$. Therefore, from $V^0(x) \leq \gamma U(x,0)$ one has $\overline{V}^0(x) \leq 2\gamma U(x,0)$. Hence, $U(x,0)/\overline{V}^0(x) \geq 1/(2\gamma), \forall x$. Therefore, if $2c/(1-c)^2 < 1/(2\gamma)$ or equivalently if
\begin{equation}
		c^2 - (2+4\gamma)c + 1 > 0, \label{AVI_Stabil_Thm_eq8_2}
\end{equation}
then inequality (\ref{AVI_Stabil_Thm_eq8}) holds.
The root of the left hand side of the foregoing inequality are real and given by
\begin{equation}
		c_1 = 1 + 2\gamma - \sqrt{4\gamma^2 + 4 \gamma} \mbox{ and } c_2 = 1 + 2\gamma + \sqrt{4\gamma^2 + 4 \gamma}.
		 \label{AVI_Stabil_Thm_eq9}
\end{equation}
Inequality (\ref{AVI_Stabil_Thm_eq8_2}) holds if $c < c_1$ or if $c > c_2$ by analysis of the sign of the quadratic equation on its left hand side. But, $c_2 > 1$, hence, any $c$ which satisfies $c > c_2$ will be unacceptable for our purpose, because, such a $c$ does not belong to $[0,1)$. As for $c < c_1$ it is required to make sure $c_1>0$, otherwise no suitable $c$ will be resulted from this analysis. From $4\gamma^2 + 4\gamma + 1 >  4\gamma^2 + 4\gamma$ which along with the non-negativeness of both sides of the last inequality leads to $\sqrt{4\gamma^2 + 4\gamma + 1} = 2\gamma+1 >  \sqrt{4\gamma^2 + 4\gamma}$, one has  $ 1 + 2\gamma - \sqrt{4\gamma^2 + 4 \gamma} > 0$. Therefore, $c_1$ is indeed positive and a non-negative $c$ smaller than $c_1$ leads to the desired stability. 
The first part of the theorem is proved by noticing that when (\ref{AVI_Stabil_Thm_eq1}) holds, $\Delta \hat{V}^i(x)$ is strictly less than zero, considering the assumed positive definiteness of $U(.,0)$. 

Finally, considering the line of proof at the end of the proof of Theorem \ref{Thm_Stab_LyapFun} it is straight forward to see that $\Delta \hat{V}^i(x) < 0$ leads to $\hat{\mathcal{B}}^{i}_{\bar{r}^i} \subset \Omega$ being an EROA for the system operated with $\hat{h}^i(.)$. The reason is, any trajectory initiated within $\hat{\mathcal{B}}^{i}_{\bar{r}^i}$ will remain inside the set and hence, within $\Omega$. Note that the continuity of the function approximators leads to the desired continuity of $\hat{V}^i(.)$ for guaranteeing the compactness of the EROA, as detailed in the proof of Theorem \ref{Thm_Stab_LyapFun}.
\qed

\begin{Thm} \label{Thm_Stabil_AVI_No_Lyap} 
Let the stabilizing approximate value iteration be implemented using a continuous function approximator with the bounded approximation error $| \epsilon^i(x)| \leq c U(x,0), \forall i \in \mathbb{N} \cup \{-1\},\forall x \in \Omega,$ for some $c\in [0,1)$. Moreover, let $\gamma \in \mathbb{R}_+$ be such that $V^0(x) \leq \gamma U(x,0), \forall x \in \Omega$. 
Then, the system operated using control policy $\hat{h}^k(.)$ at time $k$, that is, the control subject to adaptation in the approximate value iterations scheme asymptotically stabilizes the system about the origin if $c$ is such that 
\begin{equation}
		0 \leq c < 1 + 4\gamma - \sqrt{16\gamma^2 + 8 \gamma}. \label{AVI_Stabil_No_Lyap_Thm_eq1}
\end{equation}
\end{Thm}
\textit{Proof}: 
The idea for proof of this theorem is similar to that of Theorem \ref{Thm_Stabil_EVI_No_Lyap}, except that the `semi-monotonicity' feature presented in Lemma \ref{Lemma_AVI_SemiMonotonicity} will be used, instead of the monotonicity given by Lemma \ref{Lemma_NonDecreasing} in exact VI.
Eq. (\ref{AVI_3}) and Lemma \ref{Lemma_AVI_SemiMonotonicity} lead to
\begin{equation}
\begin{split}
		\hat{V}^{1}(x_0) = U\big(x_0,\hat{h}^0(x_0)\big) + \hat{V}^0\Big(f\big(x_0,\hat{h}^0(x_0)\big)\Big) + \\
		\epsilon^0(x_0) \leq \hat{V}^0(x_0) + 2c \breve{V}^0(x_0), \forall x_0 \in \Omega, \label{AVI_Stabil_No_Lyap_Thm_eq2}
\end{split}
\end{equation}
and similarly
\begin{equation}
\begin{split}
		\hat{V}^{2}(x) = U\big(x,\hat{h}^1(x)\big) + \hat{V}^1\Big(f\big(x,\hat{h}^1(x)\big)\Big) + \\
		\epsilon^1(x) \leq \hat{V}^1(x) + 2c \breve{V}^1(x) 
		, \forall x \in \Omega. \label{AVI_Stabil_No_Lyap_Thm_eq3}
\end{split}
\end{equation}
which along with (\ref{AVI_Stabil_No_Lyap_Thm_eq2}) leads to
\begin{equation}
\begin{split}
		U\big(x,\hat{h}^1(x)\big) + &\hat{V}^1\Big(f\big(x,\hat{h}^1(x)\big)\Big) + \epsilon^1(x) \leq \\
		&\hat{V}^0(x) + 2c \breve{V}^0(x) + 2c \breve{V}^1(x), \forall x \in \Omega. \label{AVI_Stabil_No_Lyap_Thm_eq3_1}
\end{split}
\end{equation}
or equivalently
\begin{equation}
\begin{split}
		U\big(x,&\hat{h}^1(x)\big) + \hat{V}^1\Big(f\big(x,\hat{h}^1(x)\big)\Big) + \\
		&\epsilon^1(x_0) - 2c \breve{V}^0(x) - 2c \breve{V}^1(x)\leq \hat{V}^0(x), \forall x \in \Omega. \label{AVI_Stabil_No_Lyap_Thm_eq3_2}
\end{split}
\end{equation}
Let $\hat{x}_k^* := f\big(\hat{x}_{k-1}^*,\hat{h}^{k-1}(\hat{x}_{k-1}^*)\big)$ for $k \geq 1$ and $\hat{x}_{0}^*:=x_0$. Evaluating (\ref{AVI_Stabil_No_Lyap_Thm_eq3_2}) at $\hat{x}_1^*$ and replacing the $V^0(\hat{x}_1^*)$ in the left hand side of the inequality in (\ref{AVI_Stabil_No_Lyap_Thm_eq2}) with the left hand side of (\ref{AVI_Stabil_No_Lyap_Thm_eq3_2}), which is smaller per (\ref{AVI_Stabil_No_Lyap_Thm_eq3_2}), one has
\begin{equation}
\begin{split}
		U\big(\hat{x}_0^*,\hat{h}^0(\hat{x}_0^*)\big) + U\big(\hat{x}_1^*,\hat{h}^1(\hat{x}_1^*)\big) + \hat{V}^1(\hat{x}_2^*) + \epsilon^0(\hat{x}_0^*) \\
		+ \epsilon^1(\hat{x}_1^*) - 2c \breve{V}^0(\hat{x}_0^*) - 2c \breve{V}^0(\hat{x}_1^*) - 2c \breve{V}^1(\hat{x}_1^*) \leq  \\
		\hat{V}^0(\hat{x}_0^*), \forall \hat{x}_0^* \in \Omega. \label{AVI_Stabil_No_Lyap_Thm_eq4}
\end{split}
\end{equation}
Repeating this process by replacing $\hat{V}^1(x_2^*)$ in (\ref{AVI_Stabil_No_Lyap_Thm_eq4}) using
\begin{equation}
\begin{split}
		U\big(x,&\hat{h}^2(x)\big) + \hat{V}^2\Big(f\big(x,\hat{h}^2(x)\big)\Big) + \epsilon^2(x) \\
		&- 2c \breve{V}^1(x) - 2c \breve{V}^2(x) \leq \hat{V}^1(x), \forall x \in \Omega, \label{AVI_Stabil_No_Lyap_Thm_eq5}
\end{split}
\end{equation}
which is resulted from 
\begin{equation}
\begin{split}
		\hat{V}^{3}(x) = &U\big(x,\hat{h}^2(x)\big) + \hat{V}^2\Big(f\big(x,\hat{h}^2(x)\big)\Big)  \\
		+\epsilon^2(x)		&\leq \hat{V}^2(x) + 2c \breve{V}^2(x) \leq  \\
		&\hat{V}^1(x) + 2c \breve{V}^1(x) + 2c \breve{V}^2(x), \forall x \in \Omega, \label{AVI_Stabil_No_Lyap_Thm_eq5_2}
\end{split}
\end{equation}
leads to
\begin{equation}
\begin{split}
		U\big(\hat{x}_0^*,\hat{h}^0(\hat{x}_0^*)\big) + U\big(\hat{x}_1^*,\hat{h}^1(\hat{x}_1^*)\big) + U\big(\hat{x}_2^*,\hat{h}^2(\hat{x}_2^*)\big) + \\
		\hat{V}^2(\hat{x}_3^*) + \epsilon^0(\hat{x}_0^*) + \epsilon^1(\hat{x}_1^*)+ \epsilon^2(\hat{x}_2^*) \\
		 - 2c \breve{V}^0(\hat{x}_0^*) - 2c \breve{V}^0(\hat{x}_1^*) - 2c \breve{V}^1(\hat{x}_1^*) - 2c \breve{V}^1(\hat{x}_2^*) \\
		- 2c \breve{V}^2(\hat{x}_2^*)\leq \hat{V}^0(\hat{x}_0^*), \forall \hat{x}_0^* \in \Omega. \label{AVI_Stabil_No_Lyap_Thm_eq6}
\end{split}
\end{equation}
Similarly by repeating this process one has
\begin{equation}
\begin{split}
		\sum_{k=0}^{i-1} \Big( U\big(\hat{x}_k^*,\hat{h}^k(\hat{x}_k^*)\big) + \epsilon^k(\hat{x}_k^*) \Big)+ \hat{V}^{i-1}(\hat{x}_i^*) \\ 
		- 2c\sum_{k=0}^{i-1} \breve{V}^k(\hat{x}_k^*) -  2c\sum_{k=1}^{i-1} \breve{V}^{k-1}(\hat{x}_k^*)
		\leq \hat{V}^0(\hat{x}_0^*), \\
		\forall \hat{x}_0^* \in \Omega, \forall i \in \mathbb{N}-\{0\}. \label{AVI_Stabil_No_Lyap_Thm_eq7}
\end{split}
\end{equation}
From (\ref{AVI_Stabil_Thm_eq5}) and  $\overline{V}^0(x) \leq 2 V^0(x) \leq 2\gamma U(x,0), \forall x \in \Omega$, one has
\begin{equation}
		\breve{V}^{i}(x) \leq \frac{2\gamma}{1-c} U(x,0), \forall x \in \Omega, \forall i \in \mathbb{N}. \label{AVI_Stabil_No_Lyap_Thm_eq7_2}
\end{equation}
Also, $(1-c) U(x,0) \leq U(x,u) + \epsilon^i(x),\forall x, \forall u, \forall i$. Therefore, from (\ref{AVI_Stabil_No_Lyap_Thm_eq7}) one has
\begin{equation}
\begin{split}
		(1&-c)\sum_{k=0}^{i-1} U\big(\hat{x}_k^*,0) + \hat{V}^{i-1}(\hat{x}_i^*) - \frac{4c\gamma}{1-c}\sum_{k=0}^{i-1} U(\hat{x}_k^*,0) \\
		&-\frac{4c\gamma}{1-c}\sum_{k=1}^{i-1} U(\hat{x}_k^*,0)
		\leq \hat{V}^0(\hat{x}_0^*)
		, \forall \hat{x}_0^* \in \Omega, \forall i \in \mathbb{N}-\{0\}. \label{AVI_Stabil_No_Lyap_Thm_eq7_3}
\end{split}
\end{equation}
The last inequality holds even if we add the negative scalar value of $-4c\gamma/(1-c)U(x_0^*,0)$ to the last summation on its left hand side, which leads to
\begin{equation}
\begin{split}
		\big((1-c)-\frac{8c\gamma}{1-c}\big)\sum_{k=0}^{i-1} U\big(\hat{x}_k^*,0) +\hat{V}^{i-1}(\hat{x}_i^*)\leq \\
		\hat{V}^0(\hat{x}_0^*)
		, \forall \hat{x}_0^* \in \Omega, \forall i \in \mathbb{N}-\{0\}. \label{AVI_Stabil_No_Lyap_Thm_eq7_4}
\end{split}
\end{equation}
Since $\hat{V}^{i-1}(x) \geq 0, \forall x$, it can be removed from the left hand side of the foregoing inequality while the inequality still holds.
Assume 
\begin{equation}
		(1-c)-\frac{8c\gamma}{1-c} > 0 \Leftrightarrow c^2 - (2+8\gamma)c+1> 0. \label{AVI_Stabil_No_Lyap_Thm_eq7_5}
\end{equation}
Comparing (\ref{AVI_Stabil_No_Lyap_Thm_eq7_4}), after removing $\hat{V}^{i-1}(x_i^*)$, with (\ref{Thm_Stab_NoLyap_7}) where the states were propagated using the exact VI in the latter, the same stability result can be obtained for the states propagated using the approximate VI, providing (\ref{AVI_Stabil_No_Lyap_Thm_eq7_5}) holds. That is, the sequence of partial sums of the left hand side of (\ref{AVI_Stabil_No_Lyap_Thm_eq7_4}) is upper bounded and because of being non-decreasing it converges, as $i\to\infty$, \cite{Rudin}. Considering the positive definiteness of $U(.,0)$, assumed in our AVI analyses, this leads to $x_i \to 0$, as long as the entire state trajectory is contained in $\Omega$.
Finally, in order to enforce (\ref{AVI_Stabil_No_Lyap_Thm_eq7_5}), one will need (\ref{AVI_Stabil_No_Lyap_Thm_eq1}). The details are identical to the last stages of the proof of Theorem \ref{Theorem_Stability_AVI}, as replacing $\gamma$ in (\ref{AVI_Stabil_Thm_eq8_2}) with $2\gamma$, the left hand sides of inequalities (\ref{AVI_Stabil_Thm_eq8_2}) and (\ref{AVI_Stabil_No_Lyap_Thm_eq7_5}) become identical. Moreover, the proof of positiveness of the right hand side of (\ref{AVI_Stabil_No_Lyap_Thm_eq1}) follows from the same argument presented in that proof. 
\qed

Finally, the last step of our analysis is presenting some results regarding the EROA for the system operated using evolving control policy during AVI.

\begin{Thm} \label{Thm_ROA_Evolving_AVI} 
Let the conditions of Theorem \ref{Thm_Stabil_AVI_No_Lyap} hold. Moreover, let $\hat{\mathcal{B}}^{i}_r:=\{ x\in\mathbb{R}^n : \hat{V}^{i}(x) \leq r \}$ and $\mathcal{B}^*_r:=\{ x\in\mathbb{R}^n : {V}^*(x) \leq r \}$ for any $r\in\mathbb{R}_+$. 
If ${\mathcal{B}}^{*}_{r} \subset \Omega$ for a given $r>0$, then compact set $\hat{\mathcal{B}}^{0}_{r/2}$ is an estimation of the region of attraction of the system operated using control policy $\hat{h}^k(.)$ at time $k$, that is, the control subject to adaptation.

\end{Thm}
\textit{Proof}:
Once the conditions of Theorem \ref{Thm_Stabil_AVI_No_Lyap} hold, from (\ref{AVI_Stabil_No_Lyap_Thm_eq7_4}) one has
\begin{equation}
		\hat{V}^{i-1}(\hat{x}_i^*)\leq \hat{V}^0(\hat{x}_0^*), \forall \hat{x}_0^* \in \Omega, \forall i \in \mathbb{N}-\{0\}, \label{AVI_Thm_ROA_Evolving_VI_eq2}
\end{equation}
where $\hat{x}_k^* := f\big(\hat{x}_{k-1}^*,\hat{h}^{k-1}(\hat{x}_{k-1}^*)\big)$ for $k \geq 1$ and $\hat{x}_{0}^*:=x_0$, that is, $\hat{x}_k^*$ denotes the state trajectory generated by the \textit{evolving} control policy from the AVI. Therefore,
\begin{equation}
x_0 \in \hat{\mathcal{B}}^{0}_{r} \Rightarrow \hat{x}_{k}^* \in \hat{\mathcal{B}}^{k-1}_{r}, \forall k \in \mathbb{N}-\{0\}, \forall r \in\mathbb{R}_+. \label{AVI_Thm_ROA_Evolving_VI_eq3}
\end{equation}
From $\underline{V}^k(x) \leq \hat{V}^k(x), \forall x, \forall k,$ cf. Theorem \ref{Thm_Boundedness}, one has 
\begin{equation}
\hat{\mathcal{B}}^{k}_{r} \subset \underline{\mathcal{B}}^{k}_{r}, \forall k \in \mathbb{N}, \forall r \in\mathbb{R}_+, \label{AVI_Thm_ROA_Evolving_VI_eq4}
\end{equation}
where $\underline{\mathcal{B}}^{i}_r:=\{ x\in\mathbb{R}^n : \underline{V}^{i}(x) \leq r \}$ and $\underline{V}^k(x)$ is defined in Theorem \ref{Thm_Boundedness}. Moreover, $\{ \underline{V}^k(x) \}_{k=0}^{\infty}$ is non-increasing and converges to $\underline{V}^*(x)$, i.e., the optimal value function corresponding to cost function (\ref{Cost_Lower_1}), because it is resulted from an exact VI. Therefore, $\underline{V}^*(x) \leq \underline{V}^{k+1}(x) \leq \underline{V}^k(x), \forall x,$ which leads to
\begin{equation}
\underline{\mathcal{B}}^{k}_{r} \subset \underline{\mathcal{B}}^{k+1}_{r} \subset \underline{\mathcal{B}}^*_{r}, \forall k \in \mathbb{N}, \forall r \in\mathbb{R}_+. \label{AVI_Thm_ROA_Evolving_VI_eq5}
\end{equation}
From (\ref{AVI_Thm_ROA_Evolving_VI_eq4}) and (\ref{AVI_Thm_ROA_Evolving_VI_eq5}) one has
\begin{equation}
\hat{\mathcal{B}}^{k}_{r} \subset \underline{\mathcal{B}}^*_{r}, \forall k \in \mathbb{N}, \forall r \in\mathbb{R}_+. \label{AVI_Thm_ROA_Evolving_VI_eq6}
\end{equation}
On the other hand, it is not hard to see that $V^*(x) \leq 2 \underline{V}^*(x), \forall x,$ which if holds leads to
\begin{equation}
\underline{\mathcal{B}}^{*}_{r} \subset \mathcal{B}^*_{2r}, \forall r \in\mathbb{R}_+. \label{AVI_Thm_ROA_Evolving_VI_eq7}
\end{equation}
To verify $V^*(x) \leq 2 \underline{V}^*(x)$, let the state trajectory generated from the exact VI of $\underline{V}^*(x)$ be denoted with $\underline{x}_k^*, \forall k$. Comparing cost function (\ref{Cost_Lower_1}) with (\ref{CostFunction}) one has
\begin{equation}
V^*(x_0) \leq \underline{V}^*(x_0) + c \sum_{k=0}^{\infty} U(\underline{x}_k^*,0), \forall x_0. \label{AVI_Thm_ROA_Evolving_VI_eq8}
\end{equation}
The reason is, both sides of the foregoing inequality are infinite sums of similar summands, except that the the left hand side is evaluated along the optimal trajectory with respect to (\ref{CostFunction}) while the right hand side is evaluated along the optimal trajectory with respect to (\ref{Cost_Lower_1}). Moreover, by composition $\sum_{k=0}^{\infty} U(\underline{x}_k^*,0) \leq \underline{V}^*(x_0)$ which along with $c <1$ and (\ref{AVI_Thm_ROA_Evolving_VI_eq8}) leads to $V^*(x) \leq 2 \underline{V}^*(x)$.

Finally, from (\ref{AVI_Thm_ROA_Evolving_VI_eq3}), (\ref{AVI_Thm_ROA_Evolving_VI_eq6}), and (\ref{AVI_Thm_ROA_Evolving_VI_eq7}) one has
\begin{equation}
x_0 \in \hat{\mathcal{B}}^{0}_{r} \Rightarrow \hat{x}_{k}^* \in {\mathcal{B}}^{*}_{2r}, \forall k \in \mathbb{N}-\{0\}, \forall r \in\mathbb{R}_+. \label{AVI_Thm_ROA_Evolving_VI_eq9}
\end{equation}
Therefore, any state trajectory initiated within $\hat{\mathcal{B}}^{0}_{r/2}$ remains inside ${\mathcal{B}}^{*}_{r}$ and if the $r$ in the latter is such that ${\mathcal{B}}^{*}_{r} \subset \Omega$, the state trajectory remains within $\Omega$ and by Theorem \ref{Thm_Stabil_AVI_No_Lyap} converges to the origin.
\qed

\section{Conclusions}
A set of theoretical analyses on convergence, stability, and regions of attraction for a value iteration based scheme which is initiated using an admissible guess were presented. Afterwards, the results were extended to the more interesting but challenging case of incorporating the approximation errors in the iterations. Simple and straight forward conditions for guaranteeing the boundedness of the learning results and the stability of the system under a fixed as well as an evolving control policy were developed. These results are expected to lay the foundation for improving the mathematical rigor of the popular field of intelligent control.

\vspace{-5pt}
\begin{wrapfigure}{l}{0.15\textwidth}
  \vspace{-15pt}
  \begin{center}
		\includegraphics[width=0.31\columnwidth]{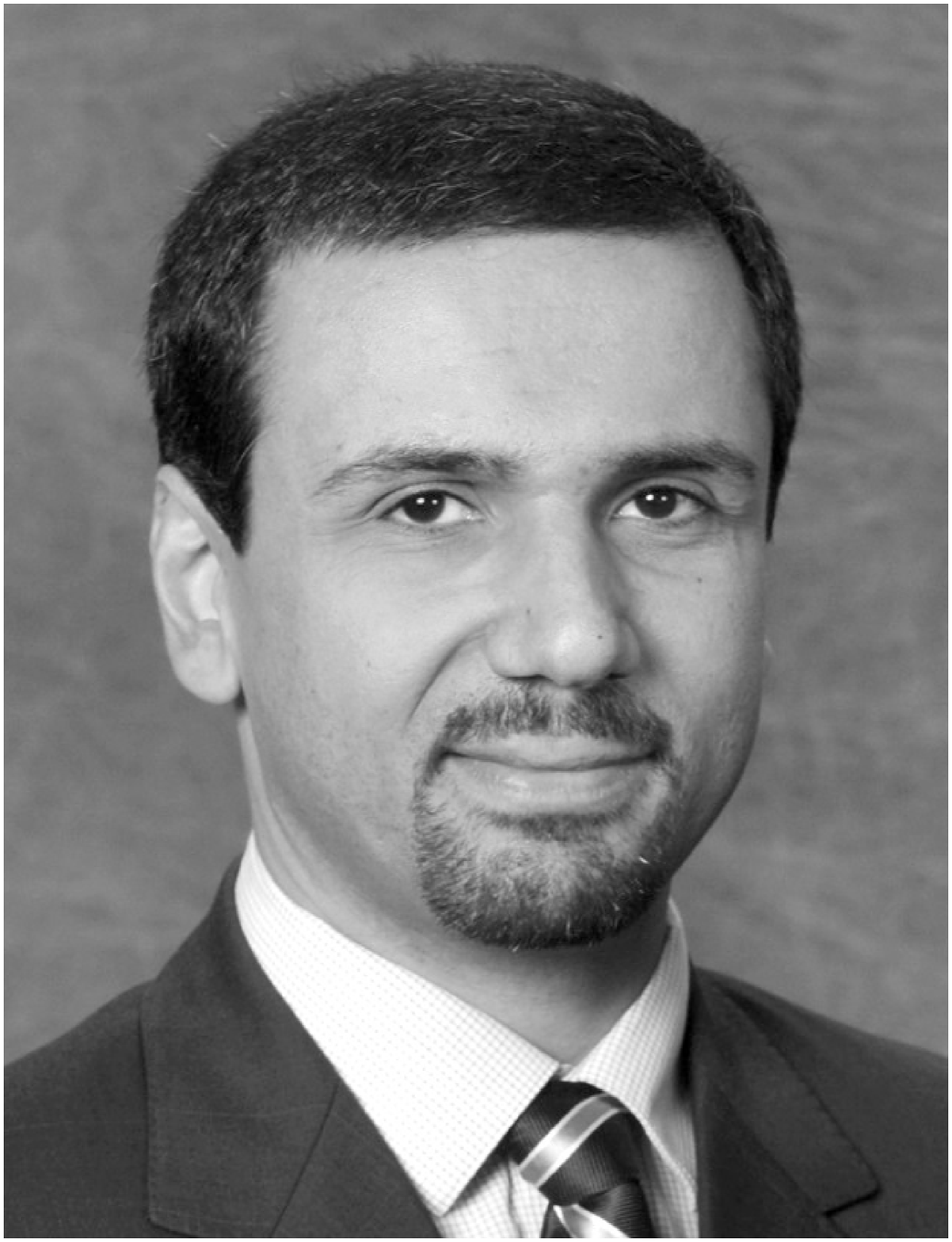}
  \end{center}
  \vspace{-0pt}
  \vspace{-15pt}
\end{wrapfigure}
\vspace{10pt}
\footnotesize
Ali Heydari received his PhD degree from the Missouri University of Science and Technology in 2013. He is currently an assistant professor of mechanical engineering at the South Dakota School of Mines and Technology. He was the recipient of the Outstanding M.Sc. Thesis Award from the Iranian Aerospace Society, the Best Student Paper Runner-Up Award from the AIAA Guidance, Navigation and Control Conference, and the Outstanding Graduate Teaching Award from the Academy of Mechanical and Aerospace Engineers at Missouri S\&T. His research interests include approximate dynamic programming and control of hybrid systems. He is a member of Tau Beta Pi.


\begin{thebibliography}{10}

\bibitem{Watkins}
C.~Watkins, {\em Learning from Delayed Rewards}.
\newblock {PhD} Dissertation, Cambridge University, Cambridge, England, 1989.

\bibitem{Werbos}
P.~J. Werbos, ``Approximate dynamic programming for real-time control and
  neural modeling,'' in {\em Handbook of Intelligent Control} (D.~A. White and
  D.~A. Sofge, eds.), Multiscience Press, 1992.

\bibitem{Sutton}
R.~S. Sutton and A.~G. Barto, {\em Reinforcement Learning: An Introduction}.
\newblock MIT Press, 1998.

\bibitem{Bertsekas_NDP}
D.~P. Bertsekas and J.~N. Tsitsiklis, {\em Neuro-Dynamic Programming}.
\newblock Athena Scientific, 1996.

\bibitem{Powell_ADP_Book}
W.~B. Powell, {\em Approximate Dynamic Programming}.
\newblock Hoboken, NJ, Wiley, 2007.

\bibitem{Bala_Biega}
S.~N. Balakrishnan and V.~Biega, ``Adaptive-critic based neural networks for
  aircraft optimal control,'' {\em Journal of Guidance, Control and Dynamics},
  vol.~19, pp.~893--898, 1996.

\bibitem{Prokhorov}
D.~Prokhorov and D.~Wunsch, ``Adaptive critic designs,'' {\em IEEE Transactions
  on Neural Networks}, vol.~8, pp.~997--1007, 1997.

\bibitem{Venaya}
G.~Venayagamoorthy, R.~Harley, and D.~Wunsch, ``Comparison of heuristic dynamic
  programming and dual heuristic programming adaptive critics for neurocontrol
  of a turbogenerator,'' {\em IEEE Transactions on Neural Networks}, vol.~13,
  pp.~764--773, May 2002.

\bibitem{Si}
R.~Enns and J.~Si, ``Helicopter trimming and tracking control using direct
  neural dynamic programming,'' {\em IEEE Transactions on Neural Networks},
  vol.~14, no.~4, pp.~929--939, 2003.

\bibitem{AlTamimi}
A.~Al-Tamimi, F.~Lewis, and M.~Abu-Khalaf, ``Discrete-time nonlinear {HJB}
  solution using approximate dynamic programming: Convergence proof,'' {\em
  IEEE Transactions on Systems, Man, and Cybernetics, Part B: Cybernetics},
  vol.~38, pp.~943--949, Aug 2008.

\bibitem{Jagannathan_Xu_Zhao}
Q.~Zhao, H.~Xu, and S.~Jagannathan, ``Optimal control of uncertain quantized
  linear discrete-time systems,'' {\em International Journal of Adaptive
  Control and Signal Processing}, 2014.

\bibitem{Liu_TCYB}
D.~Liu and Q.~Wei, ``Finite-approximation-error-based optimal control approach
  for discrete-time nonlinear systems,'' {\em IEEE Transactions on
  Cybernetics}, vol.~43, pp.~779--789, April 2013.

\bibitem{Heydari_NN_FinHor}
A.~Heydari and S.~N. Balakrishnan, ``Fixed-final-time optimal control of
  nonlinear systems with terminal constraints,'' {\em Neural Networks},
  vol.~48, pp.~61--71, 2013.

\bibitem{LewisContSystMag}
F.~Lewis, D.~Vrabie, and K.~Vamvoudakis, ``Reinforcement learning and feedback
  control: Using natural decision methods to design optimal adaptive
  controllers,'' {\em IEEE Control Systems}, vol.~32, pp.~76--105, 2012.

\bibitem{Liu_PI}
D.~Liu and Q.~Wei, ``Policy iteration adaptive dynamic programming algorithm
  for discrete-time nonlinear systems,'' {\em IEEE Transactions on Neural
  Networks and Learning Systems}, vol.~25, pp.~621--634, 2014.

\bibitem{Landelius_PhDThesis}
T.~Landelius, {\em Reinforcement learning and distributed local model
  synthesis}.
\newblock {PhD} Dissertation, Linkoping Univ., Linkoping, Sweden, 1997.

\bibitem{Bala_Liu_Convergence}
X.~Liu and S.~Balakrishnan, ``Convergence analysis of adaptive critic based
  optimal control,'' in {\em Proceedings of the American Control Conference},
  vol.~3, pp.~1929--1933, 2000.

\bibitem{Lincol_RelaxingDynProg}
B.~Lincoln and A.~Rantzer, ``Relaxing dynamic programming,'' {\em IEEE
  Transactions on Automatic Control}, vol.~51, pp.~1249--1260, Aug 2006.

\bibitem{Rinehart_VI_TAC}
M.~Rinehart, M.~Dahleh, and I.~Kolmanovsky, ``Value iteration for (switched)
  homogeneous systems,'' {\em IEEE Transactions on Automatic Control}, vol.~54,
  no.~6, pp.~1290--1294, 2009.

\bibitem{Heydari_TCYB}
A.~Heydari, ``Revisiting approximate dynamic programming and its convergence,''
  {\em IEEE Transactions on Cybernetics}, vol.~44, no.~12, pp.~2733--2743,
  2014.

\bibitem{Singh_Discounted_ADP_ErrorAnalysis}
S.~P. Singh and R.~C. Yee, ``An upper bound on the loss from approximate
  optimal-value functions,'' {\em Mach. Learn.}, vol.~16, pp.~227--233,
  1994.

\bibitem{Szepesv_AVI_API_ErrorAnalysis}
A.~Farahmand, C.~Szepesv\'{a}ri, and R.~Munos, ``Error propagation for
  approximate policy and value iteration,'' in {\em Advances in Neural
  Information Processing Systems} (J.~Lafferty, C.~Williams, J.~Shawe-Taylor,
  R.~Zemel, and A.~Culotta, eds.), pp.~568--576, 2010.

\bibitem{Szepesv_AVI_ErrorAnalysis}
R.~Munos and C.~Szepesv\'{a}ri, ``Finite-time bounds for fitted value
  iteration,'' {\em J. Mach. Learn. Res.}, vol.~9, pp.~815--857, 2008.

\bibitem{Liu_Stable_VI}
Q.~Wei and D.~Liu, ``A novel iterative $\theta$-adaptive dynamic programming
  for discrete-time nonlinear systems,'' {\em IEEE Transactions on Automation
  Science and Engineering}, vol.~11, no.~4, pp.~1176--1190, 2014.

\bibitem{Khalil}
H.~Khalil, {\em Nonlinear Systems}.
\newblock Prentice-Hall, 2002.
\newblock pp. 111-194.

\bibitem{Rudin}
W.~Rudin, {\em Principles of Mathematical Analysis}.
\newblock McGraw-Hill, 3rd~ed., 1976.
\newblock pp. 55, 60, 86, 87, 89, 145, 148.

\bibitem{Heydari_AVI}
A.~Heydari, ``Theoretical and Numerical Analysis of Approximate Dynamic Programming with Approximation Errors,'' available at arXiv.org.
	
\bibitem{Weierstrass_Theorem}
H.~Jeffreys and B.~S. Jeffreys, ``Weierstrass's theorem on approximation by
  polynomials,'' in {\em Methods of Mathematical Physics}, pp.~446--448,
  Cambridge University Press, 3rd~ed., 1988.

\bibitem{Hornik_NN_Continuity}
K.~Hornik, M.~Stinchcombe, and H.~White, ``Multilayer feedforward networks are
  universal approximators,'' {\em Neural Networks}, vol.~2, no.~5,
  pp.~359--366, 1989.

\bibitem{Kirk}
D.~E. Kirk, {\em Optimal control theory; an introduction}.
\newblock Prentice-Hall, 1970.
\newblock pp. 53-106.

\bibitem{Heydari_Franklin}
A.~Heydari and S.~Balakrishnan, ``Optimal switching between autonomous
  subsystems,'' {\em Journal of the Franklin Institute}, vol.~351, 2014.

\bibitem{Relaxing_DP_2}
A.~Rantzer, ``Relaxed dynamic programming in switching systems,'' {\em Control
  Theory and Applications, IEE Proceedings}, vol.~153, no.~5, pp.~567--574,
  2006.

\bibitem{Jadbabaie_MPC_CLF}
A.~Jadbabaie, J.~Yu, and J.~Hauser, ``Unconstrained receding-horizon control of
  nonlinear systems,'' {\em IEEE Transactions on Automatic Control}, vol.~46,
  no.~5, pp.~776--783, 2001.

\bibitem{Sontag}
E.~Sontag, ``A 'universal' construction of {A}rtstein's theorem on nonlinear
  stabilization,'' {\em System \& Control Letters}, vol.~13, pp.~117--123,
  1989.

\end{thebibliography}
\end{document}